\documentclass[journal]{IEEEtran}
\usepackage{balance}
\usepackage{amsmath,amsfonts}
\usepackage{array}
\usepackage[caption=false,font=scriptsize,labelfont=sf,textfont=sf]{subfig}
\usepackage{textcomp}
\usepackage{stfloats}
\usepackage{url}
\usepackage{graphicx}
\usepackage{cite}
\usepackage{booktabs}
\usepackage{multirow}
\usepackage{xspace}
\usepackage{textgreek}
\newcommand{\Dynawo}{Dyna\textomega o\xspace}
\newcommand{\marktext}[1]{\textcolor{black}{#1}}
\usepackage{xcolor}

\definecolor{darkgreen}{RGB}{0,128,0}
\usepackage{tikz}
\usepackage{pgfplots}
\pgfplotsset{compat=1.16}
\usetikzlibrary{positioning, arrows.meta, shapes.geometric, calc}

\makeatletter
\def\@IEEEtablecaptionsepspace{\vspace*{1pt}\relax}
\def\@IEEEfigurecaptionsepspace{\vspace*{2pt}\relax}
\makeatother

\begin{document}

\title{Graph-Based Dynamic Activity Prediction for \\ Real-Time Power System Applications}

\author{Stefanos~Eleftheriadis,
Mohammad~Hashemnezhad,
Savvas~Panagi,
Thibaut~Vermeulen,
and~Petros~Aristidou%
\thanks{This work was supported in part by the European Commission under Grant Agreements No.~101136119 (TwinEU) and No.~101120278 (DENSE), and by the TRAISIM project funded by CRESYM.}%
\thanks{S.~Eleftheriadis, M.~Hashemnezhad, S.~Panagi, and P.~Aristidou are with the Department of Electrical Engineering, Computer Engineering and Informatics, Cyprus University of Technology, 3036 Limassol, Cyprus (e-mail: \{sc.eleftheriadis, ma.hashemnezhad\}@edu.cut.ac.cy; \{savvas.panagi, petros.aristidou\}@cut.ac.cy).}%
\thanks{T.~Vermeulen is with RTE, Paris, France (e-mail: thibaut.vermeulen@rte-france.com).}}


\maketitle

\begin{abstract}
Online dynamic security assessment and day-ahead planning require rapid identification of where disturbances will produce significant dynamic activity. Full time-domain simulations are computationally demanding, while existing machine-learning approaches predict system-level outcomes or require post-disturbance trajectories. This paper proposes a pair-aware Graph Isomorphism Network with Edge features (GINE) framework comprising two task-specific models for predicting bus-voltage and generator apparent-power activity from the operating point and contingency description. Physically aware KPI cuts convert simulated responses into ordinal activity targets for model training. Evaluations on Nordic~32 and the French transmission system with more than 6\,000 buses show that pair-aware fusion and graph embeddings improve on a local baseline on Nordic~32, with GINE the lower-cost graph encoder, and that the framework remains effective at large scale. As a downstream application, the predicted activity maps guide Adaptive Model Simplification, reducing continuous and discrete model variables by 72\% and 66\%, achieving speedups of up to 3.16x, and preserving voltage trajectories when critical substations are retained. These results demonstrate scalable, pre-disturbance prediction of component-level dynamic activity.
\end{abstract}


\section{Introduction}
\label{sec:intro}

\IEEEPARstart{M}{odern} power systems operate as large cyber-physical infrastructures in which generation, wide-area measurements, and automated controls interact across multiple time scales. The dynamic response to a disturbance depends strongly on the operating point and on the contingency location, and the growing share of converter-interfaced resources reinforces this dependence. As a result, disturbances no longer produce slowly evolving, spatially uniform responses, especially in large-scale power systems. Dynamic stress can appear abruptly and concentrate in particular regions of the network, while other regions remain comparatively unaffected.

For operators, this spatial heterogeneity makes advance knowledge of the expected dynamic response indispensable. Online Dynamic Security Assessment (DSA) depends on recognizing contingency risk in time for preventive or corrective action. The same capability can also support day-ahead planning by indicating where significant dynamic behavior is expected, without requiring exhaustive simulation of all candidate scenarios. When full transient trajectories are required, real-time dynamic simulation must provide them within wall-clock budgets compatible with online decision-making. Operator training platforms~\cite{dispatcher_bucciero_1991,uses_chu_1991,experiences_miller_1993,PSCC_2026} further rely on such simulations to recreate control-room conditions with which trainees can interact as events unfold. Across these settings, speed and fidelity must be achieved together.

Exhaustive full-order simulation of every relevant contingency is, however, computationally prohibitive on large networks with hundreds of thousands of state variables. An alternative is to anticipate, from the pre-disturbance operating condition and the disturbance description alone, where post-disturbance dynamic activity will concentrate and which parts of the model must retain full detail and which admit simplification, before any time-domain simulation is carried out. Throughout this paper, dynamic activity denotes the intensity of the post-contingency response of an individual component, quantified by the Key Performance Indicators (KPIs) defined in Section~\ref{sec:methodology}, and is distinct from a stability status.

\subsection{Literature Review}

A substantial body of work aims to obtain dynamic information faster than exhaustive full-order simulation, motivated by the needs of online assessment and real-time operation. The literature has advanced along three complementary directions: data-driven prediction of post-contingency behavior, model reduction for faster simulation, and indicator-based localization of transient response.

Several studies have explored machine-learning surrogates for dynamic security assessment. Early work showed that synchronized post-fault measurements contain sufficient information for rapid transient-stability classification~\cite{support_gomez_2011}. Topology-aware deep models subsequently extended this approach by embedding network structure directly in the learning process~\cite{rotor_sun_2024}, while more recent recurrent formulations improved the treatment of sequential dynamics~\cite{grubased_okasha_2026}. At the same time, another line of research estimated stability directly from the pre-contingency state through steady-state or static features~\cite{power_an_2019,predicting_titz_2024,zhang_tsa_scrf_2026}. Although these approaches can avoid repeated time-domain simulations during online assessment, they primarily predict stability labels, margins, or trajectories rather than the spatial distribution of dynamic activity.

Model reduction addresses the same computational burden by simplifying the system model itself~\cite{aththanayake_reduction_2024}. Classical structure-preserving techniques such as coherency-based aggregation~\cite{identification_podmore_1978} and system-theoretic reductions such as Krylov-subspace methods~\cite{model_chaniotis_2005} provide rigorous simplification strategies, but are often sensitive to operating point and reduction setup. Measurement-based dynamic equivalents later improved practicality by identifying reduced behavior from boundary measurements and adapting it as conditions change~\cite{measurementbased_chai_2016,online_jiang_2021}. Nevertheless, these methods assume that the region of interest has already been identified, and accelerate simulation only after the network has been partitioned.

Complementing both directions, KPIs quantify how strongly and where the system responds after a scenario has been simulated. They support offline identification of dynamically relevant components and simulation acceleration~\cite{parallel_aristidou_2014,timevariant_xia_2022}, yet are rarely used as prediction targets for adaptive online simplification.

\subsection{Research Gap and Motivation}

Despite these advances, the spatial distribution of dynamic activity remains unavailable before a time-domain simulation is run. Graph-based learning has recently produced generator-level stability status from pre-fault or steady-state inputs~\cite{huang_gtsa_2023,huang_gtsa_2025}, identified critical generators~\cite{recurrent_huang_2020}, and screened contingencies across changing operating domains~\cite{lu_screening_2024}; these outputs indicate whether and where instability occurs, not how strongly each component responds. KPIs describe that response intensity only after trajectories are computed, and reduction methods assume the region of interest is already known. This paper therefore targets an ordinal map of component-level response intensity, defined for a given operating condition and disturbance and designed to drive adaptive simplification. Most related studies also remain limited to IEEE-scale benchmarks~\cite{recurrent_huang_2020}.

This work proposes a graph-based framework that predicts component-level dynamic activity from the pre-disturbance operating state and disturbance description, using KPI-derived labels for supervision. Adaptive model simplification for real-time simulation is demonstrated as a downstream use of the predicted activity map; online DSA, day-ahead planning, and operator training are prospective applications that are not evaluated in this work.

The remainder of the paper is organized as follows. Section~\ref{sec:background} summarizes the graph-based learning background and the GINE formulation underlying the proposed approach. Section~\ref{sec:methodology} presents the methodology. Section~\ref{sec:case_study} describes the Nordic~32 and French-system case studies. Section~\ref{sec:results} reports the numerical results. Section~\ref{sec:discussion} discusses the findings, and Section~\ref{sec:conclusion} concludes the paper.

\section{Background}
\label{sec:background}

This section reviews the graph representation of the power system and the Graph Isomorphism Network with Edge features (GINE) underlying the proposed model.

\subsection{Graph Representation of the Power System}

The power system is represented by a graph $\mathcal{G}=(\mathcal{V},\mathcal{E})$, whose nodes are electrical components and whose edges are the interconnections between them. Node $i\in\mathcal{V}$ carries a feature vector $\mathbf{x}_i\in\mathbb{R}^{F_n}$, and edge $(i,j)\in\mathcal{E}$ an attribute vector $\mathbf{e}_{ij}\in\mathbb{R}^{F_e}$. The neighbors of node $i$ are $\mathcal{N}(i)=\{j\in\mathcal{V}\mid (i,j)\in\mathcal{E}\}$. Message passing forms a hidden state $\mathbf{h}_i^{(l)}\in\mathbb{R}^{H}$ by combining the state of node $i$ with aggregated information from $\mathcal{N}(i)$~\cite{xu2019howpowerful}, starting from $\mathbf{h}_i^{(0)}=\psi(\mathbf{x}_i)$.

\subsection{Graph Isomorphism Network with Edge Features}

GINE extends the Graph Isomorphism Network~\cite{xu2019howpowerful} by letting edge attributes shape the messages between connected nodes~\cite{hu2020strategies}. In a power system those attributes describe the line, transformer, or event associated with a connection, so the exchanged message depends on more than topology~\cite{varbella2024powergraph,kazim2025edgegnn}. Node and edge features are mapped once into a common hidden dimension $H$; the edge map is $\tilde{\mathbf{e}}_{ij}=\psi_e(\mathbf{e}_{ij})$. With activation $\phi(\cdot)$ (commonly $\mathrm{ReLU}$), learnable scalar $\epsilon^{(l)}$, and multilayer perceptron $\mathrm{MLP}^{(l)}(\cdot)$, each layer updates the node state by
\begin{equation}
\mathbf{h}_{i}^{(l+1)}=\mathrm{MLP}^{(l)}\!\Big((1+\epsilon^{(l)})\mathbf{h}_{i}^{(l)}+\textstyle\sum_{j\in\mathcal{N}(i)}\phi\!\big(\mathbf{h}_{j}^{(l)}+\tilde{\mathbf{e}}_{ij}\big)\Big),
\label{eq:gine_update}
\end{equation}
aggregating the edge-conditioned messages from $\mathcal{N}(i)$.

\section{Methodology}
\label{sec:methodology}

The proposed framework is organized into two stages: offline learning and online inference. During offline learning, post-contingency simulations are represented as power-system graphs, component-level KPIs are extracted, and physically aware cuts convert them into dynamic-activity targets for supervised learning. The simulations run for $100$\,s with maximum step size $h_{\max}=1$\,s, and the contingency is applied at $t=10$\,s. During online inference, the trained framework uses the operating point and contingency description to estimate where dynamic activity is likely to occur before a time-domain simulation is performed. The resulting component-level activity map can then support different downstream applications, such as adaptive model simplification.

\subsection{Power-System Graph Construction}

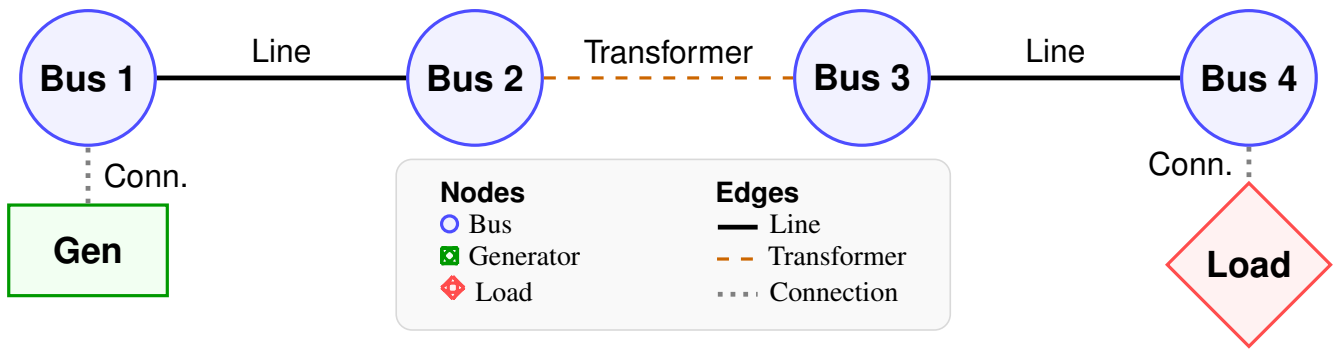
\begin{figure}
\centering
\resizebox{\columnwidth}{!}{\begin{tikzpicture}[
    node distance=1.5cm and 2.2cm,
    busbar/.style={
        circle,
        draw=blue!70,
        fill=blue!5,
        thick,
        minimum size=11mm,
        font=\small\sffamily\bfseries
    },
    generator/.style={
        rectangle,
        draw=green!60!black,
        fill=green!5,
        thick,
        minimum width=14mm,
        minimum height=8mm,
        font=\small\sffamily\bfseries
    },
    load/.style={
        diamond,
        draw=red!70,
        fill=red!5,
        thick,
        minimum size=10mm,
        font=\small\sffamily\bfseries
    },
    line/.style={draw=black, very thick},
    transformer/.style={draw=orange!80!black, thick, dashed},
    connection/.style={draw=gray, very thick, dotted}
]

    \node[busbar] (bb1) {Bus 1};
    \node[generator, below=of bb1, yshift=10mm] (gen) {Gen};

    \node[busbar, right=of bb1] (bb2) {Bus 2};
    \node[busbar, right=of bb2] (bb3) {Bus 3};
    \node[busbar, right=of bb3] (bb4) {Bus 4};

    \node[load, below=of bb4, yshift=12mm] (load) {Load};

    \draw[connection]
        (gen) -- node[right, font=\footnotesize\sffamily] {Conn.} (bb1);

    \draw[line]
        (bb1) -- node[above, font=\footnotesize\sffamily] {Line} (bb2);

    \draw[transformer]
        (bb2) -- node[above, font=\footnotesize\sffamily] {Transformer} (bb3);

    \draw[line]
        (bb3) -- node[above, font=\footnotesize\sffamily] {Line} (bb4);

    \draw[connection]
        (bb4) -- node[left, font=\footnotesize\sffamily] {Conn.} (load);

    \node[
        draw=gray!30,
        fill=gray!5,
        rounded corners,
        inner sep=5pt,
        font=\scriptsize,
        right=2cm of gen,
        yshift=0.5mm
    ] (legend) {
        \begin{tabular}{lll}
            \textbf{\sffamily Nodes} & & \textbf{\sffamily Edges} \\

            {\tikz\draw[blue!70, fill=blue!5, thick] (0,0) circle (2pt);}
            Bus
            &
            \hspace{0.35cm}
            &
            \tikz[baseline=-0.5ex]\draw[line] (0,0) -- (10pt,0);
            Line \\

            {\tikz\draw[green!60!black, fill=green!5, thick] (0,0) rectangle (4pt,4pt);}
            Generator
            &
            &
            \tikz[baseline=-0.5ex]\draw[transformer] (0,0) -- (10pt,0);
            Transformer \\

            {\tikz\draw[red!70, fill=red!5, thick,rotate=45] (0,0) rectangle (4pt,4pt) ;}
            Load
            &
            &
            \tikz[baseline=-0.5ex]\draw[connection] (0,0) -- (10pt,0);
            Connection
        \end{tabular}
    };

\end{tikzpicture}}
\caption{Graph construction principle used to map the power-system model into a homogeneous graph representation.}
\label{fig:graph_construction}
\end{figure}

Each graph $\mathcal{G}=(\mathcal{V},\mathcal{E})$ is one operating-point--contingency scenario, with buses, generators, and loads as nodes and lines, transformers, and connection links as edges~(Fig.~\ref{fig:graph_construction}). A homogeneous representation is adopted to keep message passing and optimization simple and stable. Although heterogeneous architectures can define separate transformations for every component and relation type, they also introduce additional parameters and model complexity. In the adopted representation, node and edge types are encoded directly as input features alongside the electrical quantities, allowing the model to distinguish component roles while retaining one shared message-passing scheme. Under contingency~$c$, the features of node~$i$ are
$\mathbf{x}_i=\bigl[t_i,\,v_i,\,\theta_i,\,p_i,\,q_i,\,f_i,\,d_{i,c}\bigr],$
where $t_i$ denotes the node type, $v_i$ the steady-state voltage in~kV, $\theta_i$ the angle, $(p_i,q_i)$ the injections in~MW/Mvar, $f_i$ a binary fault indicator, and $d_{i,c}$ the log-transformed transfer-impedance magnitude to the contingency.

The transfer-impedance magnitude is computed from the bus-admittance matrix $\mathbf{Y}$. The bus-impedance matrix is obtained as $\mathbf{Z}=\mathbf{Y}^{-1}$, and the transfer-impedance magnitude between buses $i$ and $j$ is defined as $d_{ij}=\lvert Z_{ij}\rvert$. For a node contingency at bus $a$, the magnitude of node $i$ to the contingency is $\bar{d}_{i,c}=d_{ia}$. For an edge contingency between buses $u$ and $v$, it is defined as the magnitude to the nearest terminal, $\bar{d}_{i,c}=\min(d_{iu},d_{iv})$. The logarithmic transformation $d_{i,c}=\log(1+\bar{d}_{i,c})$ is applied to reduce the skewness of the transfer-impedance-magnitude distribution and limit the influence of large values~\cite{best_west_2021}. The transfer-impedance magnitude $\lvert Z_{ij}\rvert$ indicates the electrical coupling between bus $i$ and the disturbance location $j$. It is not an electrical distance. It tells the model how strongly a disturbance at $j$ can influence $i$, in addition to the graph topology. For edge $(i,j)$, the features are
$\mathbf{e}_{ij}=\bigl[c_{ij},\,f_{ij},\,r_{ij},\,x_{ij},\,b_{1,ij},\,g_{1,ij},\,b_{2,ij},\,g_{2,ij}\bigr],$
where $c_{ij}$ denotes the edge type, $f_{ij}$ the fault flag, and the remaining entries are series and shunt parameters ($r,x$ in~$\Omega$; $b,g$ in~S).

\subsection{KPI Extraction}

Dynamic activity is quantified from post-contingency trajectories of the full-detail model by component-level KPIs~\cite{parallel_aristidou_2014}: voltage magnitude (pu) on buses and $S/S_{\mathrm{nom}}$ on generators (pu). In both cases the KPI is the maximum variance over a moving window of length $\Delta T_{\mathrm{w}}$, advanced by $\Delta T_{\mathrm{s}}$,
\begin{equation}
\begin{aligned}
\chi_i^{(k)}
&=
\mathrm{Var}\!\bigl(y_i(t)\bigr),
\qquad
t\in[t_k,t_k+\Delta T_{\mathrm{w}}],
\\
t_{k+1}
&=
t_k+\Delta T_{\mathrm{s}},
\qquad
\chi_i
=
\max_k \chi_i^{(k)},
\end{aligned}
\label{eq:kpi_maxvar}
\end{equation}
where $y_i(t)$ is the relevant trajectory. For the case studies in this paper, $\Delta T_{\mathrm{w}}=5$\,s, $\Delta T_{\mathrm{s}}=1$\,s, and the first window opens at $t_0=9$\,s. Each window variance is the mean squared deviation from the sample mean, using the recorded samples with no resampling. Samples that are not finite are dropped. A trajectory that is empty, identically zero, or shorter than one window receives no KPI. From $t=10$\,s onward, components that undergo disconnection, de-energization, or a control action, including a protection trip, are identified on the simulated response and assigned a non-KPI flag class~$C-1$ (non-simplifiable); they are excluded from the cut-based label construction of the following subsection. That flag class is learned rather than set by a handcrafted rule. At prediction time only the pre-disturbance operating point and the contingency description are available as inputs.

\subsection{Physically Aware Cut Selection and Label Construction}

Scenarios are split at the operating-point level into train, validation, and test sets, so all contingencies of one operating point stay together. Cut selection and feature scaling use training statistics only; the test set therefore evaluates unseen operating points. Component and contingency identities recur across the split, since both are embedded from a fixed vocabulary; the method therefore addresses interpolation across operating points of a fixed network, and components or contingencies outside that vocabulary are not handled by the present formulation.

Raw KPIs are continuous and skewed, so they are mapped to ordered severity labels. Let $x\ge 0$ be a KPI and $\boldsymbol{\tau}=(\tau_0,\ldots,\tau_{B-1})$ an ordered cut set with $0<\tau_0<\cdots<\tau_{B-1}$. The ordinal class is
\begin{equation}
c(x;\boldsymbol{\tau})=\sum_{m=0}^{B-1}\mathbb{I}(x>\tau_m),
\end{equation}
with $B$ cuts, $c(x;\boldsymbol{\tau})\in\{0,\ldots,B\}$; together with the flag class, the model output therefore has $C=B+2$ classes. In this work $B=4$ and $C=6$: classes $0$ to $4$ are KPI classes and class~$5$ is the flag~$C-1$.
With solver precision $\nu$, a trajectory within numerical tolerance yields variance at most $\nu^{2}$; the inactive cut is therefore $\tau_0=\alpha\,\nu^{2}$ with margin $\alpha\ge 1$.

The remaining cuts are selected on training scenarios by a multi-objective procedure~\cite{deb2002nsga2,marquez2020emodits}. Each candidate keeps $\tau_0$ fixed and sets $(\tau_1,\tau_2,\tau_3)=A(\beta_1,\beta_2,\beta_3)$, where the scale and the three fractions are taken from
\begin{equation}
\begin{aligned}
A&\in\{5,7.5,10,15,20,30,40,50,75,100\}\times 10^{-5},\\
\beta_1&\in\{0.005,0.01,0.02,0.05,0.075,0.10\},\\
\beta_2&\in\{0.03,0.05,0.075,0.10,0.15,0.25,0.40\},\\
\beta_3&\in\{0.15,0.20,0.30,0.40,0.50,0.60,0.75\}.
\end{aligned}
\label{eq:cut_grid}
\end{equation}
A combination is kept only when $0<\beta_1<\beta_2<\beta_3<1$ and $\tau_0<\tau_1<\tau_2<\tau_3$. The grid contains $1\,890$ ladders. Each candidate is scored on \emph{usability} (learnable class shares), \emph{physical consistency} (higher classes match stronger post-event behavior), and \emph{scenario-level coherence} (more severe scenarios receive higher aggregate classes).

Let $\mathbf{p}=(p_1,\ldots,p_K)$ be the shares of classes $1$ to $K$ among components whose KPI exceeds the inactive cut, renormalized so that these shares sum to one. Class~$0$ is excluded, and $K=4$. Usability combines a normalized entropy of shares,
\begin{equation}
H(\mathbf{p})
=
-\frac{1}{\log K}\sum_{j=1}^{K} p_j\log p_j,
\end{equation}
a class-floor term that rewards minimum shares $p_2^{\min}=0.02$, $p_3^{\min}=0.005$, and $p_4^{\min}=0.001$ for the stronger KPI classes,
\begin{equation}
U_{\mathrm{floor}}(\mathbf{p})
=
\min_{j\in\{2,\ldots,K\}}
\min\!\left(1,\frac{p_j}{p_j^{\min}}\right),
\end{equation}
and a soft penalty that discourages a weakest active class larger than $p_1^{\max}=0.92$,
\begin{equation}
U_{1}(\mathbf{p})
=
\begin{cases}
1, & p_1\le p_1^{\max},\\
\min\!\left(1,\dfrac{1-p_1}{1-p_1^{\max}}\right), & \text{otherwise}.
\end{cases}
\end{equation}
The usability score is then
\begin{equation}
U
=
0.45\,H(\mathbf{p})
+0.35\,U_{\mathrm{floor}}(\mathbf{p})
+0.20\,U_{1}(\mathbf{p}).
\label{eq:usability}
\end{equation}
Physical consistency checks that higher activity classes match more severe post-event waveforms. A morphological severity index $\varphi$ is evaluated on training components above the inactive cut. In this work, about $400$ trajectories per task group are used. 
Because the values are strongly skewed toward inactivity, a uniform draw would over-represent mild responses; the set is therefore stratified by $\log_{10}$ decade (each integer power of ten), sampling evenly across nonempty decades with a fixed seed. For each selected trajectory,
\begin{equation}
\varphi=
\log_{10}\!\left(
\eta + E + \sqrt{\max(V,0)} + \lambda_D D
\right),
\label{eq:phi}
\end{equation}
where $E$ is the largest absolute deviation from the pre-event mean during the $50$~s after the disturbance, $V$ is the variance of that interval, and $D$ is the mean absolute deviation over its last $20\%$. The three terms use the units of the recorded trajectory. The constants are $\eta=10^{-12}$ and $\lambda_D=0.5$. Given ordinal KPI activity classes $c$,
\begin{equation}
P
=
0.65\,\max\bigl(\rho(\varphi,c),0\bigr)
+0.35\,M(\varphi,c),
\label{eq:physical}
\end{equation}
where $\rho$ is the Spearman correlation of $\varphi$ and $c$ on average ranks, and $M$ is the fraction of adjacent classes that are both nonempty and whose mean $\varphi$ does not decrease. Empty classes are omitted. If no adjacent pair remains, $M$ contributes $0$. If fewer than five finite pairs remain, or either ranking does not vary, $\rho$ is undefined and contributes $0$. Scenario coherence is the Spearman correlation $R$ between the 95th-percentile KPI of a scenario and the mean activity class of its components. The percentile marks the severe tail of the scenario, and the mean class checks whether that severity appears in the labels. The combined score is
\begin{equation}
J
=
0.40\,U
+0.40\,P
+0.20\,\max(R,0).
\label{eq:combined_cut}
\end{equation}
Candidates that are not dominated in $(U,P)$ form a Pareto set. The labeling cut is the feasible candidate with largest $J$. Feasibility requires the active-class share constraints of~\eqref{eq:usability}, namely $p_1\le p_1^{\max}$ and $p_j\ge p_j^{\min}$ for $j\in\{2,3,4\}$, together with $P\ge 0.15$ in~\eqref{eq:physical}.

\subsection{Pair-Aware GINE Training}
\label{sec:pairaware_gine}

\begin{figure*}[t]
\centering
\includegraphics[width=\textwidth]{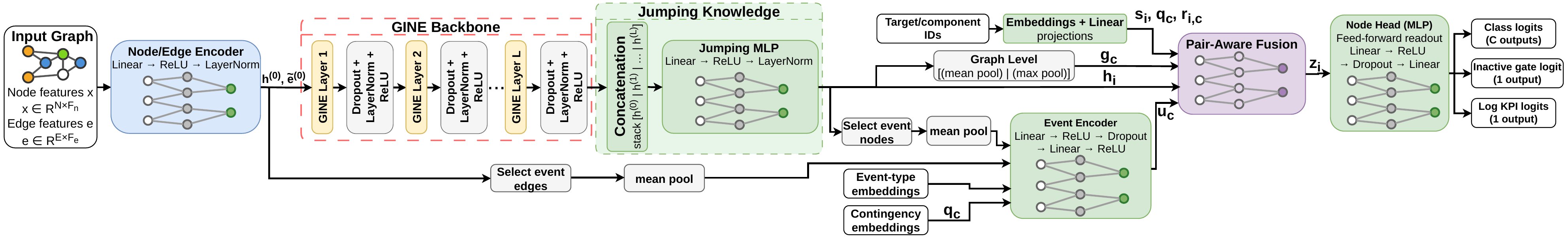}
\vspace{-5mm}
\makeatletter
\begingroup
\renewcommand{\@makecaption}[2]{%
  \@IEEEfigurecaptionsepspace
  \par\centering\normalfont\footnotesize {#1.}\nobreakspace\nobreakspace #2\par
}
\caption{Architecture of the proposed pair-aware GINE model for component-level dynamic activity prediction.}
\label{fig:pairaware_gine_arch}
\endgroup
\makeatother
\end{figure*}

Two independent GINE-based predictors are trained: a bus-level voltage-activity model and a generator-level apparent-power-activity model. Although they target different physical quantities, both use the same graph representation and edge-aware message-passing architecture. The voltage model is supervised only on bus nodes, whereas the apparent-power model is supervised only on generator nodes, restricting each model to the relevant component type. Their shared architecture is shown in Fig.~\ref{fig:pairaware_gine_arch}. The MLP topology is kept fixed, with Linear--ReLU--LayerNorm encoders, a two-layer $H\!\to\!2H\!\to\!H$ MLP in each GINE block, and a two-layer shared decoder. The hidden dimension $H$, type, identity, and pair-embedding dimensions, number of GINE layers $L$, decoder width, and dropout rate are treated as model hyperparameters.

\paragraph{Encoder and GINE backbone}
Continuous features are concatenated with learnable type and identity embeddings and mapped into width~$H$. Residual GINE layers~\cite{hu2020strategies,kazim2025edgegnn} update node states via~\eqref{eq:gine_update}. After $L$ residual GINE layers, jumping-knowledge concatenation combines the initial and intermediate node states $[\mathbf{h}_i^{(0)}|\cdots|\mathbf{h}_i^{(L)}]$; a Linear--ReLU--LayerNorm block then maps the concatenated representation to the final node embedding $\mathbf{h}_i$.

\paragraph{Event, graph, and pair context}
Contingency $c$ is encoded as a node or edge event using Boolean selection masks. A node event selects the disturbed node, while an edge event selects the faulted branch and its terminal nodes. The selected final node embeddings and encoded edge embeddings are mean-pooled separately; the edge-pooled vector is zero for a node event. The two pooled vectors are concatenated with the event-type embedding and contingency-identity embedding and passed through a small MLP to obtain $\mathbf{u}_c=\Gamma_{\mathrm{event}}(\mathcal{G},c)$.
A graph-level context is obtained by concatenating mean and max pooling over all node embeddings,
$\mathbf{g}_c=\bigl[\mathrm{mean}(\{\mathbf{h}_j\}),\;\mathrm{max}(\{\mathbf{h}_j\})\bigr].$
Target and contingency identity embeddings $\mathbf{s}_i$ and $\mathbf{q}_c$ interact as
$\mathbf{r}_{i,c}=\tanh\!\bigl(\mathbf{W}_s\mathbf{s}_i\odot\mathbf{W}_q\mathbf{q}_c\bigr),$
so the same component can receive different predictions under different contingencies. 

\paragraph{Pair-aware fusion and three decoding paths}
Interaction features between the target embedding and the event context are collected as
\begin{equation}
\boldsymbol{\delta}_{i,c}=
\bigl[
\mathbf{h}_i,\;
\mathbf{u}_c,\;
\mathbf{g}_c,\;
|\mathbf{h}_i-\tilde{\mathbf{u}}_c|,\;
\mathbf{h}_i\odot\tilde{\mathbf{u}}_c,\;
d_{i,c}
\bigr],
\label{eq:pair_delta}
\end{equation}
where $\tilde{\mathbf{u}}_c$ is a linear projection of $\mathbf{u}_c$ onto the node embedding space and $d_{i,c}$ is the transfer-impedance magnitude from graph construction. The decoder input is then
$\mathbf{z}_{i,c}=
\bigl[\boldsymbol{\delta}_{i,c},\;\mathbf{s}_i,\;\mathbf{q}_c,\;\mathbf{r}_{i,c}\bigr].$
A shared MLP maps $\mathbf{z}_{i,c}$ to three paths aligned with Fig.~\ref{fig:pairaware_gine_arch}:
\begin{equation}
\boldsymbol{\ell}_{i,c}=g_{\mathrm{cls}}(\mathbf{z}_{i,c}),\;
a_{i,c}=g_{\mathrm{gate}}(\mathbf{z}_{i,c}),\;
\hat{z}_{i,c}=g_{\mathrm{reg}}(\mathbf{z}_{i,c}),
\end{equation}
namely $C$-way class logits (activity classes and flag class~$C-1$), an inactivity-gate logit, and a standardized log-KPI estimate. All three paths are trained jointly; one decode path is selected at deployment.

\paragraph{Supervision and model selection}
Let $y_{i,c}\in\{0,\ldots,C-1\}$ with KPI classes $0,\ldots,C-2$ and flag~$C-1$. The gate target is $\mathbb{I}(y_{i,c}=0)$. Where a finite KPI exists, the regression target is the train-set standardized value
\begin{equation}
z_{i,c}^{\star}=\bigl(\log_{10}(\chi_i+\varepsilon)-\mu\bigr)/\sigma,
\end{equation}
with training mean $\mu$ and standard deviation $\sigma$. Training minimizes
\begin{equation}
\mathcal{L}
=
w_{\mathrm{cls}}\mathcal{L}_{\mathrm{CE}}
+w_{\mathrm{reg}}\mathcal{L}_{\mathrm{Huber}}
+w_{\mathrm{gate}}\mathcal{L}_{\mathrm{BCE}}
+w_{\mathrm{ord}}\mathcal{L}_{\mathrm{CDF}},
\end{equation}
with class weights $\omega_k=\sqrt{N/\max(n_k,1)}$, where $n_k$ is the number of supervised samples in class $k$ and $N$ is their total, then divided by the mean of $\{\omega_k\}$ so that the weights have unit mean, and gate positive weight $\omega_{+}=n_{\neq 0}/\max(n_{0},1)$. Each term below is averaged over the batch: cross-entropy, the gate, and the CDF term over every supervised sample, and the Huber term over the finite-KPI samples only. For one sample,
\begin{equation}
\begin{aligned}
\mathcal{L}_{\mathrm{CE}}
&=
-\sum_{k=0}^{C-1}\omega_k\,\mathbb{I}(y_{i,c}=k)\log p_{i,c}^{(k)},
\\
\mathcal{L}_{\mathrm{BCE}}
&=
-\omega_{+}\,y_{0}\log\sigma(a_{i,c})
-(1-y_{0})\log\!\bigl(1-\sigma(a_{i,c})\bigr),
\\
\mathcal{L}_{\mathrm{Huber}}
&=
\begin{cases}
\tfrac12 r^{2}, & |r|<1,\\
|r|-\tfrac12, & |r|\ge 1,
\end{cases}
\\
\mathcal{L}_{\mathrm{CDF}}
&=
\frac{1}{C-1}\sum_{k=0}^{C-2}
\bigl|F_{i,c}(k)-F^{\star}_{i,c}(k)\bigr|,
\end{aligned}
\label{eq:loss_terms}
\end{equation}
where $p_{i,c}=\mathrm{softmax}(\boldsymbol{\ell}_{i,c})$, $y_{0}=\mathbb{I}(y_{i,c}=0)$, $r=\hat{z}_{i,c}-z_{i,c}^{\star}$, and $F_{i,c}(k)=\sum_{j=0}^{k}p_{i,c}^{(j)}$, with $F^{\star}$ the same sum on the one-hot label. Cross-entropy, the gate, and the CDF term include class~$0$ and the flag. The Huber term includes class~$0$ and excludes the flag. For the final fit, the class weights and $\omega_{+}$ are recomputed on the merged train and validation set. The cuts and $(\mu,\sigma)$ stay those of the training set and do not use the test set.
Models are optimized independently with AdamW; hyperparameters are selected with Optuna (TPE)~\cite{optuna_akiba_2019}. After each validation epoch, predictions are decoded with all three paths and scored by
\begin{equation}
\begin{aligned}
\mathrm{Score}
&=
0.4\,\mathrm{BA}
+0.3\,F_{1}^{\mathrm{m}}
+0.2\,\mathrm{Acc}
+0.1\,\mathrm{Acc}_{1},
\end{aligned}
\end{equation}
using balanced accuracy ($\mathrm{BA}$), macro-$F_1$ ($F_{1}^{\mathrm{m}}$), exact accuracy ($\mathrm{Acc}$), and within-one accuracy ($\mathrm{Acc}_{1}$). The best epoch is retained per trial. Although all three decode paths are scored, the class path is used for every result reported in Section~\ref{sec:results}; the gated and log-KPI paths are kept as alternatives. The winning trial is finally retrained on train+validation.

\subsection{Inference}
\label{sec:inference}

At inference, one operating-point--contingency graph is prepared as in training and the forward pass yields $\boldsymbol{\ell}_{i,c}$, $a_{i,c}$, and $\hat{z}_{i,c}$. One decode mode is applied:
\begin{itemize}
  \item \textbf{Class:} $\hat{y}_{i,c}=\arg\max_{k\in\{0,\ldots,C-1\}}\ell_{i,c}^{(k)}$. The flag is returned when that index is $C-1$.
  \item \textbf{Gate:} if $\sigma(a_{i,c})\ge\gamma$, return class~$0$, including when the class argmax is the flag. Otherwise $\hat{y}_{i,c}=\arg\max_{k\in\{1,\ldots,C-1\}}\ell_{i,c}^{(k)}$, so the flag can be selected and class~$0$ cannot.
  \item \textbf{Log-KPI:} denormalize $\hat{z}_{i,c}$, map it through $\boldsymbol{\tau}$ to $\{0,\ldots,C-2\}$, then replace that class by the flag if $\arg\max_k\ell_{i,c}^{(k)}=C-1$. The gate is not used.
\end{itemize}

\subsection{Adaptive Model Simplification (AMS)}
\label{sec:ams}

The predicted activity classes from the previous subsection can be used online for Adaptive Model Simplification (AMS), without a time-domain simulation. Let $\hat{y}_i^{(\mathrm{V})}$ and $\hat{y}_j^{(\mathrm{S})}$ be the decoded classes of bus~$i$ and generator~$j$, and $\mathcal{B}_s$, $\mathcal{G}_s$ the buses and dynamically modeled generators of substation~$s$. Given threshold $\kappa$, substation~$s$ retains full node-breaker detail if any component satisfies $\hat{y}_i^{(\mathrm{V})}\ge\kappa$ or $\hat{y}_j^{(\mathrm{S})}\ge\kappa$; otherwise it is collapsed to bus-breaker form. Because the flag class $C-1$ is the highest class, a flagged bus or generator forces retention for every $\kappa$. The rule takes the maximum over all components of the substation, so a low voltage class on one bus cannot override a high apparent-power class on a generator of the same substation.

\subsection{Evaluation of Adaptive Model Simplification}
\label{sec:ams_eval}

Simplified models are assessed against the Full node-breaker reference under the same simulation setup introduced at the beginning of this section, with respect to:
\begin{itemize}
  \item \textbf{Solving time:} the interval $t<9$\,s is treated as initialization and excluded. Over the remaining window $t\ge 9$\,s, $\lVert T_{\mathrm{sol}}\rVert_{\infty}$ is the largest step solve time and RT Viol.\ the number of steps whose solve time exceeds $h_{\max}$. A run is real time when RT Viol.\ $=0$, equivalently $\lVert T_{\mathrm{sol}}\rVert_{\infty}\le h_{\max}$.
  \item \textbf{Voltage fidelity:} reported as a DTW RMSE. Dynamic Time Warping (DTW) is implemented with the Python library \texttt{fastdtw}~\cite{salvador2007toward} to align each Full and simplified bus-voltage curve; the root-mean-square error (RMSE) of the aligned samples is then computed per bus. Because \texttt{fastdtw} does not normalize the curves, each DTW RMSE remains in pu, and the reported value is the maximum error over all buses.
\end{itemize}

\section{Case Study}
\label{sec:case_study}

All simulations use Dynawaltz from the \Dynawo suite~\cite{dynawo}. Two networks of complementary scale are considered. Nordic~32 has 74 buses, 52 lines, 50 two-winding transformers, 26 substations, 22 loads, and 20 synchronous generators with detailed excitation, AVR, PSS, turbine, and governor models (15--400\,kV), totaling 3\,790 continuous variables, 3\,005 discrete variables, and 1\,804 root functions.
The French RTE system has more than 6\,000 buses, 7\,075 branches, 3\,039 dynamically modeled loads, 4\,009 substations, and 839 LTCs over 2\,138 transformers (63--400\,kV). Averaged over snapshots it comprises about 370 detailed synchronous generators and about 340 simplified or renewable dynamic generator models, on which $S_{\text{power}}$ is supervised when present; the full-detail model has 374\,989 continuous, 397\,301 discrete, and 353\,160 root functions.

Datasets are independently simulated operating-point--contingency scenarios. Nordic~32 has 7\,419 scenarios from 35 operating points obtained by randomized global, regional, and per-load scaling of a common reference following Hagmar \textit{et al.}~\cite{hagmar_dvsm_2020} (Table~\ref{tab:nordic_op_generation}); the French system has 69\,616 scenarios from 61 operating points (Table~\ref{tab:rte_operating_points}). Train/validation/test splits are 25/5/5 and 49/6/6 as in Section~\ref{sec:methodology}, shared by the voltage and $S_{\text{power}}$ models.
Each scenario applies one N-1 or bus disconnection if the element is connected. On the French system, lines and buses follow an RTE N-1 list; generator trips take, per operating point, the top 10\% by generation among dynamically modeled units. Only successful runs are kept, and curves with both extrema within $10^{-12}$ of zero are discarded. Unique counts distinct contingency definitions; Retained is the dataset size. Nordic~32 is used for the full prediction pipeline and the French system for large-scale scalability and AMS. Nordic training and inference used a dual-socket AMD EPYC~7452 with 128~GiB RAM and one NVIDIA Tesla~T4 (Ubuntu~24.04); architecture comparisons and French training and evaluation used dual-socket AMD EPYC~9333 nodes with 1\,032~GB RAM and NVIDIA RTX PRO 6000 Blackwell GPUs; AMS timing used an AMD Ryzen~7 5800X with 16\,GB RAM.

\begin{table}
  \centering
  \caption{Nordic~32 operating points (OPs) and contingencies.}
  \label{tab:nordic_op_generation}
  \setlength{\tabcolsep}{3pt}
  \begin{tabular}{@{}p{0.26\columnwidth}p{0.34\columnwidth}rr@{}}
    \toprule
    \multicolumn{4}{@{}l}{\textbf{Operating points}} \\
    \midrule
    \textbf{Item} & \multicolumn{3}{l}{\textbf{Description}} \\
    \midrule
    Base OP & \multicolumn{3}{p{0.68\columnwidth}}{Total active load $\approx 11\,060$\,MW.} \\
    Loading bands & \multicolumn{3}{p{0.68\columnwidth}}{Global factor $g$: low $[0.80,0.85]$, medium $[0.86,0.90]$, high $[0.91,0.95]$.} \\
    Area patterns & \multicolumn{3}{p{0.68\columnwidth}}{\textit{uniform}; \textit{central\_up}; \textit{north\_up}; \textit{south\_up}; \textit{north\_central\_stress}; \textit{mixed}.} \\
    Per-load factor & \multicolumn{3}{p{0.68\columnwidth}}{$f_i\sim\mathcal{U}(0.80,1.20)$, power factor preserved.} \\
    Resulting OPs & \multicolumn{3}{p{0.68\columnwidth}}{35 settled snapshots covering low/medium/high loading and regional stress.} \\
    \midrule
    \multicolumn{4}{@{}l}{\textbf{Contingencies}} \\
    \midrule
    \textbf{Event} & \textbf{Action at $t=10$\,s} & \textbf{Unique} & \textbf{Retained} \\
    \midrule
    Line disc. & Both ends opened & 52 & 1\,817 \\
    Transformer disc. & Both ends opened & 50 & 1\,729 \\
    Bus disc. & Bus opened & 74 & 2\,424 \\
    Load disc. & Load opened & 22 & 768 \\
    Generator trip & Switch-off signal & 20 & 681 \\
    \midrule
    \multicolumn{2}{@{}l}{Total} & 218 & 7\,419 \\
    \bottomrule
  \end{tabular}
\end{table}

\begin{table}
  \centering
  \caption{French-system operating points (OPs) and contingencies.}
  \label{tab:rte_operating_points}
  \setlength{\tabcolsep}{3pt}
  \begin{tabular}{@{}p{0.26\columnwidth}p{0.34\columnwidth}rr@{}}
    \toprule
    \multicolumn{4}{@{}l}{\textbf{Operating points}} \\
    \midrule
    \textbf{OP} & \multicolumn{3}{l}{\textbf{Description}} \\
    \midrule
    1 & \multicolumn{3}{p{0.68\columnwidth}}{SCADA snapshot (peak load, $\sim$80\,GW).} \\
    2 & \multicolumn{3}{p{0.68\columnwidth}}{SCADA snapshot (stressed condition, France--Spain interface).} \\
    3 & \multicolumn{3}{p{0.68\columnwidth}}{SCADA snapshot (minimum load, $\sim$30\,GW).} \\
    4--5 & \multicolumn{3}{p{0.68\columnwidth}}{Variants of OP~1 via regional load scaling.} \\
    6--10 & \multicolumn{3}{p{0.68\columnwidth}}{Variants of OP~2 via uniform and regional scaling.} \\
    11 & \multicolumn{3}{p{0.68\columnwidth}}{Winter worst-case loading.} \\
    12--35 & \multicolumn{3}{p{0.68\columnwidth}}{Full summer day (hourly).} \\
    36--37 & \multicolumn{3}{p{0.68\columnwidth}}{High renewable-penetration snapshots ($\sim$40\% wind/solar).} \\
    38--61 & \multicolumn{3}{p{0.68\columnwidth}}{Full winter day (hourly).} \\
    \midrule
    \multicolumn{4}{@{}l}{\textbf{Contingencies}} \\
    \midrule
    \textbf{Event} & \textbf{Action at $t=10$\,s} & \textbf{Unique} & \textbf{Retained} \\
    \midrule
    Line disc. & Both ends opened & 860 & 48\,738 \\
    Bus disc. & Bus opened & 283 & 16\,580 \\
    Generator trip & Switch-off signal & 4\,300 & 4\,298 \\
    \midrule
    \multicolumn{2}{@{}l}{Total} & 5\,443 & 69\,616 \\
    \bottomrule
  \end{tabular}
\end{table}

\section{Results}
\label{sec:results}

This section reports cut selection, Nordic~32 tuning and architecture comparison, French-system scalability, and adaptive simplification.

\subsection{Physically Aware Cut Selection and Class Construction}
\label{sec:results_cuts}

\begin{figure}
\centering
\includegraphics[width=0.92\columnwidth]{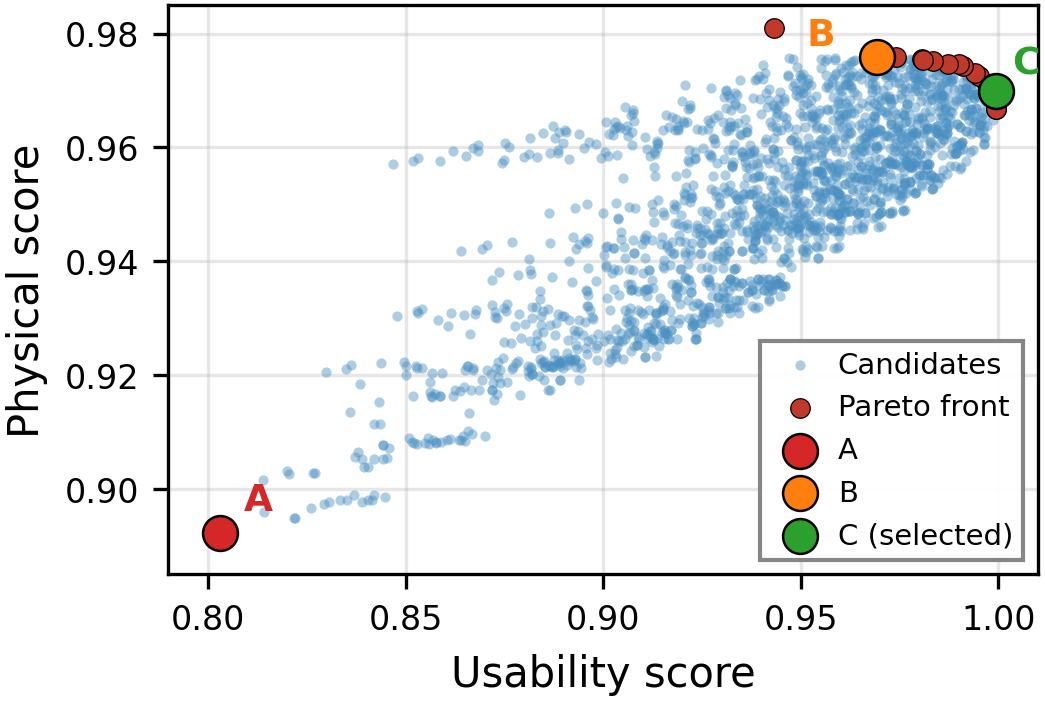}%
\vspace{-0.4\baselineskip}%
\caption{Voltage cut selection on Nordic~32 (training operating points).}
\label{fig:cut_pareto_voltage}
\end{figure}

\begin{figure}
\centering
\resizebox{0.9\columnwidth}{!}{%
\definecolor{c0}{RGB}{31,119,180}
\definecolor{c1}{RGB}{255,127,14}
\definecolor{c2}{RGB}{44,160,44}
\definecolor{c3}{RGB}{214,39,40}
\definecolor{c4}{RGB}{148,103,189}
\definecolor{event}{RGB}{80,80,80}

\begin{tikzpicture}

\begin{axis}[
  name=top,
  scale only axis,
  width=11.6cm,
  height=4.2cm,
  xmin=0, xmax=100,
  ymin=-0.11, ymax=0.05,
  ylabel={$\Delta$ (S/$S_{\mathrm{nom}}$) (pu)},
  ylabel style={font=\large},
  tick label style={font=\normalsize},
  ytick={-0.10,-0.05,0,0.025},
  yticklabels={-0.10,-0.05,0,0.025},
  scaled y ticks=false,
  grid=major,
  major grid style={line width=.15pt,draw=gray!30},
  tick style={black},
  axis y line*=left,
  axis x line*=bottom,
  xtick={0,20,40,60,80,100},
  xticklabels={,,,,,}, 
  unbounded coords=discard,
  clip=true,
]

\addplot[line width=1.25pt, color=c0]
  table [x=time, y=class_0, col sep=comma] {figures/CSV/spower_delta_example23.csv};
\addplot[line width=1.25pt, color=c1]
  table [x=time, y=class_1, col sep=comma] {figures/CSV/spower_delta_example23.csv};
\addplot[line width=1.25pt, color=c2]
  table [x=time, y=class_2, col sep=comma] {figures/CSV/spower_delta_example23.csv};
\addplot[line width=1.25pt, color=c3]
  table [x=time, y=class_3, col sep=comma] {figures/CSV/spower_delta_example23.csv};
\addplot[line width=1.25pt, color=c4]
  table [x=time, y=class_4, col sep=comma] {figures/CSV/spower_delta_example23.csv};

\addplot[dashed, color=event, line width=1.1pt, forget plot]
  coordinates {(10,-0.11) (10,0.05)};

\end{axis}

\begin{axis}[
  name=bottom,
  at={(top.south west)},
  anchor=north west,
  yshift=-0.50cm,
  scale only axis,
  width=11.6cm,
  height=4.2cm,
  xmin=0, xmax=100,
  ymin=-0.02, ymax=0.045,
  xlabel={Time (s)},
  ylabel={$\Delta$ Voltage (pu)},
  xlabel style={font=\large, yshift=0.15cm},
  ylabel style={font=\large},
  tick label style={font=\normalsize},
  ytick={-0.02,0,0.02,0.04},
  yticklabels={-0.02,0,0.02,0.04},
  scaled y ticks=false,
  grid=major,
  major grid style={line width=.15pt,draw=gray!30},
  tick style={black},
  axis y line*=left,
  axis x line*=bottom,
  xtick={0,20,40,60,80,100},
  unbounded coords=discard,
  clip=true,
]

\addplot[line width=1.25pt, color=c0]
  table [x=time, y=class_0, col sep=comma] {figures/CSV/voltage_delta_example32.csv};
\addplot[line width=1.25pt, color=c1]
  table [x=time, y=class_1, col sep=comma] {figures/CSV/voltage_delta_example32.csv};
\addplot[line width=1.25pt, color=c2]
  table [x=time, y=class_2, col sep=comma] {figures/CSV/voltage_delta_example32.csv};
\addplot[line width=1.25pt, color=c3]
  table [x=time, y=class_3, col sep=comma] {figures/CSV/voltage_delta_example32.csv};
\addplot[line width=1.25pt, color=c4]
  table [x=time, y=class_4, col sep=comma] {figures/CSV/voltage_delta_example32.csv};

\addplot[dashed, color=event, line width=1.1pt, forget plot]
  coordinates {(10,-0.02) (10,0.045)};

\end{axis}

\coordinate (legL) at ([xshift=-1.35cm]top.north west);
\coordinate (legR) at ([xshift=1.35cm]top.north east);
\node[anchor=south, font=\normalsize, inner sep=0pt] at
  ($ (legL)!0.5!(legR) + (0,0.24cm) $) {%
  \tikz[baseline=-0.5ex]\draw[line width=1.25pt,color=c0](0,0)--(0.45,0);~class 0\quad
  \tikz[baseline=-0.5ex]\draw[line width=1.25pt,color=c1](0,0)--(0.45,0);~class 1\quad
  \tikz[baseline=-0.5ex]\draw[line width=1.25pt,color=c2](0,0)--(0.45,0);~class 2\quad
  \tikz[baseline=-0.5ex]\draw[line width=1.25pt,color=c3](0,0)--(0.45,0);~class 3\quad
  \tikz[baseline=-0.5ex]\draw[line width=1.25pt,color=c4](0,0)--(0.45,0);~class 4%
};

\pgfresetboundingbox
\path
  ([xshift=-1.35cm,yshift=-0.62cm]bottom.south west)
  rectangle
  ([xshift=1.35cm,yshift=0.62cm]top.north east);

\end{tikzpicture}%
} 
\caption{Representative $\Delta$ trajectories for classes 0--4. Top: generator $S/S_{\mathrm{nom}}$. Bottom: bus voltage.}
\label{fig:delta_class_examples}
\end{figure}
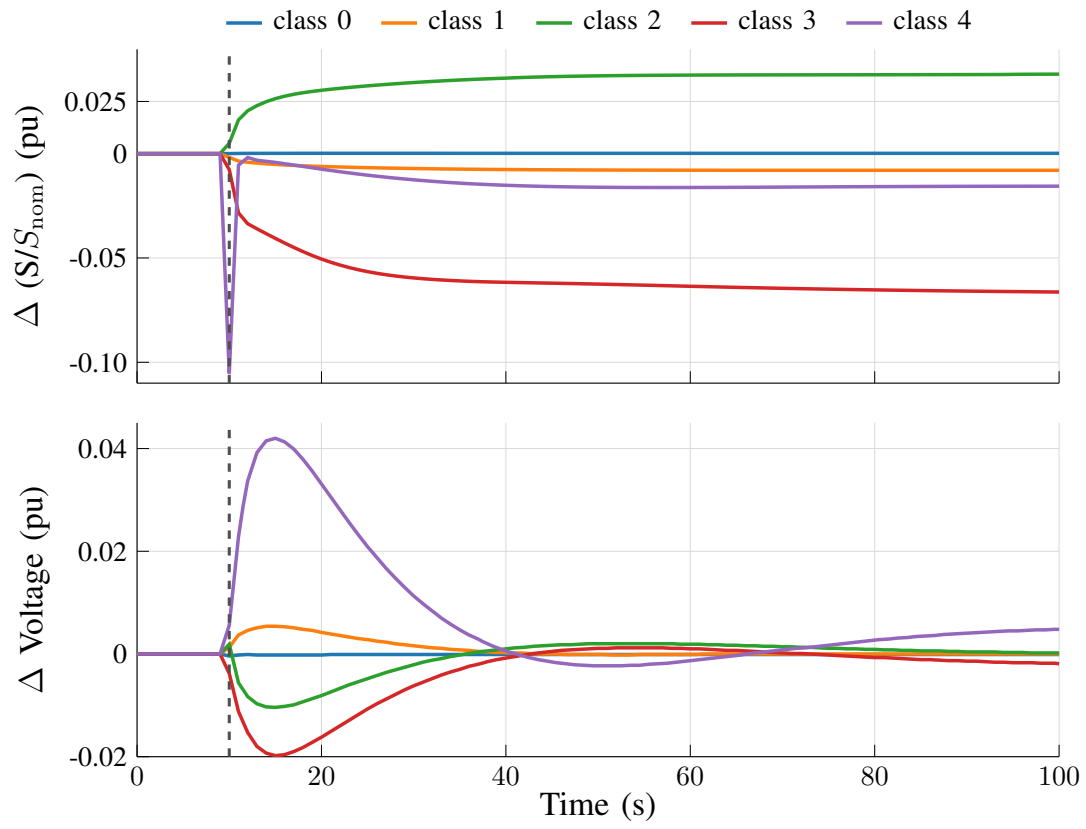

\begin{figure}
\centering
\includegraphics[width=\columnwidth]{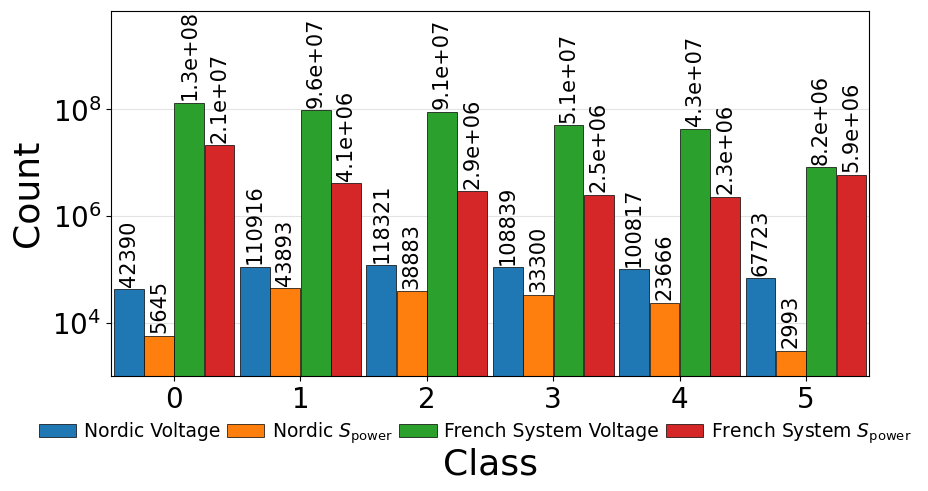}%
\vspace{-0.5\baselineskip}%
\caption{Class distributions of the voltage and apparent-power datasets after physically aware cut selection.}
\label{fig:dataset_class_dist}
\end{figure}

With $\nu=10^{-4}$ and $\alpha=10$, the inactive threshold is $\tau_0=10^{-7}$; the higher cuts follow the procedure of Section~\ref{sec:methodology} on the training operating points. On Nordic~32 voltage, $1\,890$ candidate ladders are evaluated. Table~\ref{tab:nordic_cut_candidates} lists three of them; candidate~C is selected. Fig.~\ref{fig:cut_pareto_voltage} shows the full set and the Pareto front. The same procedure is applied to apparent power and, independently, to the French system; the selected cuts are reported in Table~\ref{tab:kpi_cuts}. Fig.~\ref{fig:delta_class_examples} shows representative trajectories after centering by the pre-event mean, $\Delta(t)=y(t)-\bar{y}_{\mathrm{pre}}$, and Fig.~\ref{fig:dataset_class_dist} the resulting class counts.

\begin{table}
  \centering
  \caption{Examples of Nordic~32 voltage cut candidates.}
  \label{tab:nordic_cut_candidates}
  \setlength{\tabcolsep}{3.5pt}
  \begin{tabular}{lccc}
    \toprule
    \textbf{Quantity} & \textbf{A} & \textbf{B} & \textbf{C (selected)} \\
    \midrule
    $\tau_1$ & $3.75\times10^{-6}$ & $2.5\times10^{-6}$ & $5\times10^{-6}$ \\
    $\tau_2$ & $5\times10^{-6}$ & $3.75\times10^{-5}$ & $3.75\times10^{-5}$ \\
    $\tau_3$ & $7.5\times10^{-6}$ & $3.75\times10^{-4}$ & $1.5\times10^{-4}$ \\
    \midrule
    Class-1 share & 0.227 & 0.191 & 0.251 \\
    Class-2 share & 0.024 & 0.330 & 0.270 \\
    Class-3 share & 0.033 & 0.369 & 0.249 \\
    Class-4 share & 0.716 & 0.109 & 0.229 \\
    \midrule
    Usability score & 0.803 & 0.969 & 0.999 \\
    Physical score & 0.892 & 0.976 & 0.970 \\
    Combined score & 0.838 & 0.974 & 0.983 \\
    \bottomrule
  \end{tabular}
\end{table}

\begin{table}
  \centering
  \caption{Selected KPI cuts for Nordic~32 and the French system.}
  \label{tab:kpi_cuts}
  \setlength{\tabcolsep}{2.8pt}
  \begin{tabular}{llcccc}
    \toprule
    \textbf{System} & \textbf{Task} & $\tau_0$ & $\tau_1$ & $\tau_2$ & $\tau_3$ \\
    \midrule
    Nordic~32 & Voltage & $10^{-7}$ & $5\times10^{-6}$ & $3.75\times10^{-5}$ & $1.5\times10^{-4}$ \\
    Nordic~32 & $S_{\text{power}}$  & $10^{-7}$ & $2\times10^{-5}$ & $10^{-4}$ & $6\times10^{-4}$ \\
    French & Voltage & $10^{-7}$ & $5\times10^{-7}$ & $2.5\times10^{-6}$ & $10^{-5}$ \\
    French & $S_{\text{power}}$ & $10^{-7}$ & $5\times10^{-7}$ & $2.5\times10^{-6}$ & $7.5\times10^{-6}$ \\
    \bottomrule
  \end{tabular}
\end{table}

\subsection{Hyperparameter Optimization on Nordic~32}
\label{sec:results_hpo}
Voltage and apparent-power models are tuned independently on Nordic~32 under a shared Optuna protocol (same seed and search space). Training uses batch size~$16$, $100$~epochs with early-stopping patience~$12$, loss weights $(w_{\mathrm{cls}},w_{\mathrm{reg}},w_{\mathrm{gate}},w_{\mathrm{ord}})=(1.0,0.3,0.2,0.1)$, $\gamma=0.5$, $\varepsilon=10^{-10}$, and a fixed class decode path. Fifteen trials explore hidden width $\{128,256\}$, node-identity $\{16,24,32\}$, contingency-identity $\{16,32,64\}$, type embedding $\{8,16\}$, pair context $\{8,16,32\}$, GINE depth $\{3,4,5\}$, decoder width $\{128,256,512\}$, dropout in $[0.02,0.25]$, and log-uniform learning rate and weight decay in $[10^{-5},1.5\times10^{-3}]$ and $[10^{-7},10^{-3}]$. Table~\ref{tab:nordic_hpo} lists three candidates; candidate~C is selected for both tasks. 

\begin{table}
  \centering
  \caption{Examples of hyperparameter candidates on Nordic~32.}
  \label{tab:nordic_hpo}
  \setlength{\tabcolsep}{3.5pt}
  \begin{tabular}{lccc}
    \toprule
    \textbf{Hyperparameter} & \textbf{A} & \textbf{B} & \textbf{C (selected)} \\
    \midrule
    Hidden width & 128 & 256 & 128 \\
    Node-ID dim.\ & 16 & 32 & 32 \\
    Contingency-ID dim.\ & 32 & 64 & 16 \\
    Type dim.\ & 8 & 8 & 8 \\
    Pair dim.\ & 32 & 32 & 8 \\
    GINE layers & 4 & 5 & 5 \\
    Decoder width & 128 & 256 & 128 \\
    Dropout & 0.118 & 0.063 & 0.042 \\
    Learning rate & $1.14\times10^{-5}$ & $8.76\times10^{-4}$ & $3.08\times10^{-4}$ \\
    Weight decay & $2.70\times10^{-7}$ & $1.44\times10^{-5}$ & $5.76\times10^{-6}$ \\
    \midrule
    Best epoch (voltage) & 99 & 54 & 90 \\
    Val.\ score (voltage) & 0.859 & 0.896 & 0.906 \\
    Best epoch ($S_{\text{power}}$) & 100 & 55 & 90 \\
    Val.\ score ($S_{\text{power}}$) & 0.915 & 0.946 & 0.951 \\
    \bottomrule
  \end{tabular}
\end{table}

\subsection{Performance Comparison on Nordic~32}
\label{sec:results_arch}

\begin{figure}
\centering
\includegraphics[width=\columnwidth]{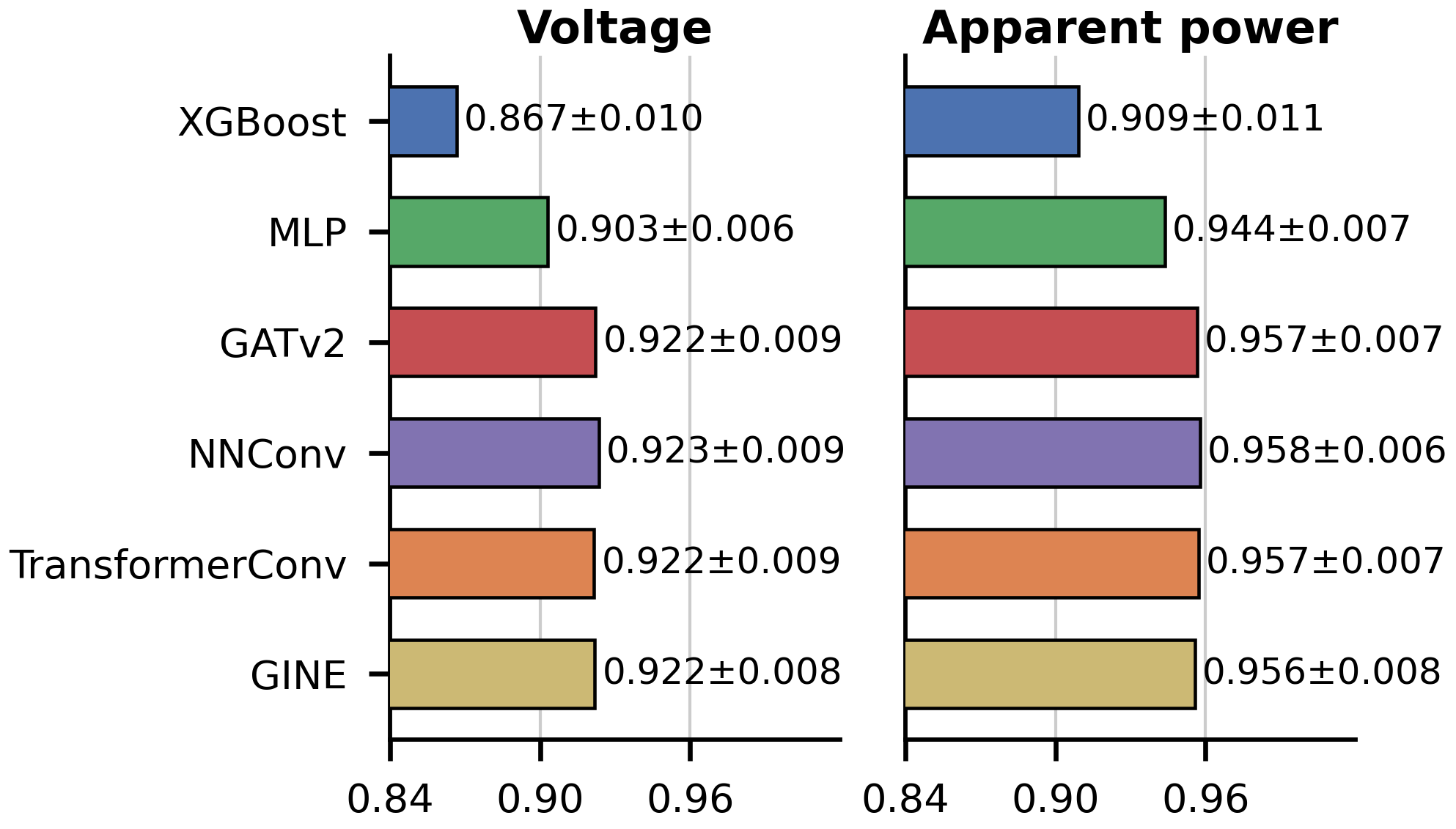}%
\vspace{-0.4\baselineskip}%
\caption{Validation selection score on Nordic~32 (5-fold mean$\pm$std).}
\label{fig:nordic_arch_val_score}
\end{figure}

Six predictors of increasing structural content are evaluated on the Nordic~32 voltage and apparent-power tasks. XGBoost~\cite{chen2016xgboost} uses only local pre-event features of each supervised target. The remaining models share the pair-aware decoder of Section~\ref{sec:methodology}. The MLP omits message passing and $\mathbf{g}_c$; its fusion is \mbox{$\boldsymbol{\delta}_{i,c}^{\mathrm{MLP}}=[\mathbf{h}_i^{(0)},\mathbf{u}_c,|\mathbf{h}_i^{(0)}-\tilde{\mathbf{u}}_c|,\mathbf{h}_i^{(0)}\odot\tilde{\mathbf{u}}_c,d_{i,c}]$}. 
GATv2~\cite{brody2021attentive}, NNConv~\cite{gilmer2017neural}, TransformerConv\cite{shi2020masked}, and GINE restore~\eqref{eq:pair_delta} with edge attributes in the convolution and $\mathbf{h}_i$ in place of $\mathbf{h}_i^{(0)}$, so only the encoder differs. For Table~\ref{tab:nordic_arch_test}, each architecture is independently tuned for each task using 5-fold cross-validation over the training and validation operating points, separate from the single-split selection of candidate~C in Table~\ref{tab:nordic_hpo}. Each Optuna trial uses the same hyperparameters across folds and is scored by the mean fold validation selection score. The winning configuration is retrained on all training and validation operating points and evaluated on the fixed held-out test set. Fig.~\ref{fig:nordic_arch_val_score} reports the mean and standard deviation of the fold validation scores for the selected trials.

\begin{table*}
  \centering
  \caption{Nordic~32 held-out test comparison. Scen.\ $\overline{F_1}$/P10~$F_1$: mean/10th percentile of scenario macro-$F_1$ over classes present in each scenario's ground truth. Cl.~5~$F_1$: pooled $F_1$ for flag class~5. Inference: median (mean$\pm$std), ms/scenario. Train wall: Optuna plus retraining.}
  \label{tab:nordic_arch_test}
  \setlength{\tabcolsep}{2.2pt}
  \footnotesize
  \resizebox{\textwidth}{!}{%
  \begin{tabular}{llcccccccclr}
    \toprule
    \textbf{Task} & \textbf{Model} & \textbf{BA} & \textbf{Acc.} & \textbf{Acc$_1$} & \textbf{Macro-$F_1$} & \textbf{Scen.\ $\overline{F_1}$} & \textbf{P10 $F_1$} & \textbf{Cl.~5 $F_1$} & \textbf{\#Params} & \textbf{Train wall} & \textbf{Infer.(ms/scenario)} \\
    \midrule
    Voltage
      & XGBoost & 0.859 & 0.852 & 0.975 & 0.858 & 0.803 & 0.580 & 0.859 & --- & 15m 19s & $0.552\,(0.552{\pm}0.004)$ \\
      & MLP & 0.893 & 0.888 & 0.977 & 0.892 & 0.855 & 0.678 & 0.871 & $6.48\times10^{5}$ & 13h 39m & $1.286\,(1.482{\pm}0.273)$ \\
      & GATv2 & 0.919 & 0.917 & 0.980 & 0.917 & 0.887 & 0.728 & 0.876 & $1.99\times10^{6}$ & 17h 15m & $1.457\,(1.642{\pm}0.285)$ \\
      & NNConv & 0.918 & 0.916 & 0.979 & 0.917 & 0.884 & 0.721 & 0.870 & $5.15\times10^{7}$ & 17h 02m & $2.058\,(2.248{\pm}0.294)$ \\
      & TransformerConv & 0.918 & 0.916 & 0.978 & 0.917 & 0.885 & 0.721 & 0.873 & $2.58\times10^{6}$ & 13h 40m & $1.424\,(1.617{\pm}0.300)$ \\
      & GINE & 0.914 & 0.913 & 0.979 & 0.914 & 0.882 & 0.728 & 0.876 & $1.60\times10^{6}$ & 16h 38m & $1.402\,(1.597{\pm}0.303)$ \\
    \midrule
    $S_{\text{power}}$
      & XGBoost & 0.883 & 0.869 & 0.989 & 0.888 & 0.830 & 0.600 & 0.997 & --- & 6m 39s & $0.161\,(0.162{\pm}0.005)$ \\
      & MLP & 0.921 & 0.908 & 0.997 & 0.920 & 0.886 & 0.685 & 0.990 & $4.79\times10^{5}$ & 5h 02m & $1.344\,(1.554{\pm}0.291)$ \\
      & GATv2 & 0.937 & 0.925 & 0.999 & 0.937 & 0.904 & 0.729 & 0.992 & $3.74\times10^{6}$ & 11h 26m & $1.497\,(1.703{\pm}0.290)$ \\
      & NNConv & 0.940 & 0.932 & 1.000 & 0.942 & 0.913 & 0.740 & 0.991 & $5.15\times10^{7}$ & 22h 59m & $2.043\,(2.236{\pm}0.299)$ \\
      & TransformerConv & 0.937 & 0.929 & 0.999 & 0.937 & 0.910 & 0.733 & 0.993 & $4.12\times10^{6}$ & 12h 36m & $1.457\,(1.684{\pm}0.314)$ \\
      & GINE & 0.934 & 0.922 & 0.998 & 0.934 & 0.898 & 0.700 & 0.984 & $2.26\times10^{6}$ & 9h 58m & $1.399\,(1.595{\pm}0.306)$ \\
    \bottomrule
  \end{tabular}%
  }
\end{table*}

\subsection{Scalability on the French Transmission System}
\label{sec:results_rte_scalability}
Relative to the Nordic~32 protocol, the French runs keep the same batch size, epoch budget, early-stopping patience, gate threshold, and class decode path. Optuna uses $48$~trials; loss weights become $(w_{\mathrm{cls}},w_{\mathrm{reg}},w_{\mathrm{gate}},w_{\mathrm{ord}})=(1.0,0.3,0.15,0.15)$; and the search widens to hidden width $\{128,256,512\}$, node-identity $\{16,24,32,64\}$, type embedding $\{8,16,32,64\}$, and pair context $\{8,16,32,64\}$. With the cuts of Table~\ref{tab:kpi_cuts}, the selected voltage model uses hidden width~$256$, embeddings $(24,32,16,32)$ for node, contingency, type, and pair dimensions, four GINE layers, and decoder width~$512$; the apparent-power model uses hidden width~$512$, embeddings $(64,32,32,8)$ in the same order, four GINE layers, and decoder width~$128$. 
Held-out metrics are in Table~\ref{tab:french_arch_test}.

\begin{table}
  \centering
  \caption{Held-out test metrics of the pair-aware GINE on the French system. Metric definitions follow Table~\ref{tab:nordic_arch_test}.}
  \label{tab:french_arch_test}
  \setlength{\tabcolsep}{2.2pt}
  \footnotesize
  \resizebox{\columnwidth}{!}{%
  \begin{tabular}{lccccccc}
    \toprule
    \textbf{Task} & \textbf{BA} & \textbf{Acc.} & \textbf{Acc$_1$} & \textbf{Macro-$F_1$} & \textbf{Scen.\ $\overline{F_1}$} & \textbf{P10 $F_1$} & \textbf{Cl.~5 $F_1$} \\
    \midrule
    Voltage & 0.700 & 0.679 & 0.930 & 0.675 & 0.608 & 0.398 & 0.715 \\
    $S_{\text{power}}$ & 0.717 & 0.792 & 0.917 & 0.714 & 0.593 & 0.451 & 0.704 \\
    \bottomrule
  \end{tabular}%
  }
\end{table}

\begin{figure}[!t]
\centering
\begin{tikzpicture}
\node (main) at (-0.45,0) {
    \includegraphics[width=0.98\columnwidth]{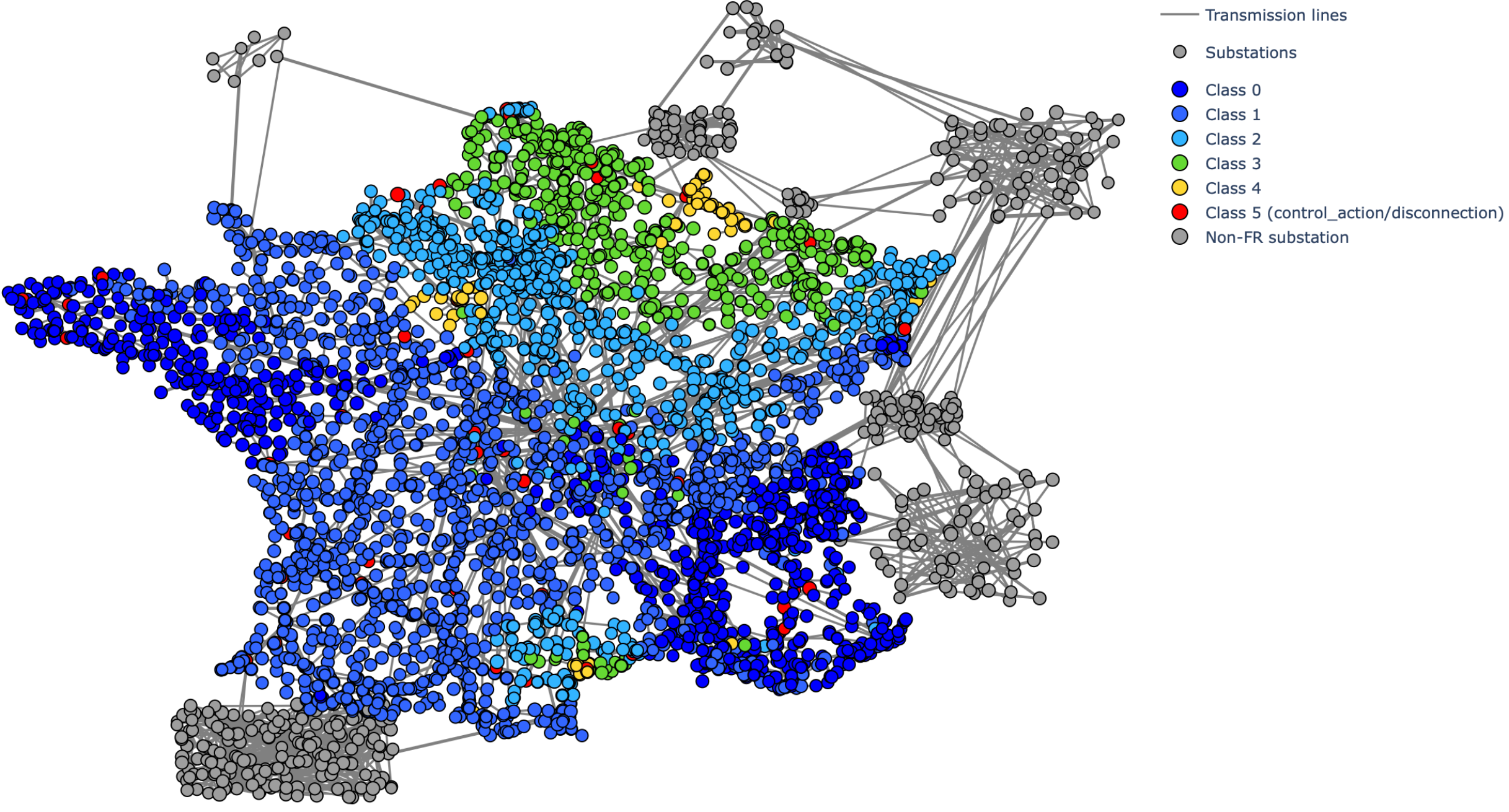}
};
\node[circle, draw=red, very thick, minimum size=7.5mm] (target) at (-0.85,1.15) {};
\node[draw, rectangle, inner sep=1.2pt] (predimg) at (-3.42,1.82) {
    \includegraphics[width=0.155\columnwidth]{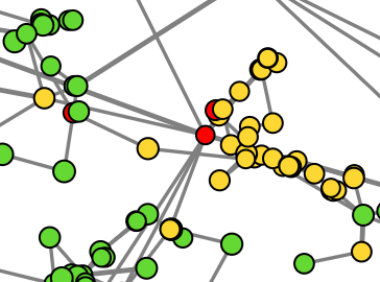}
};
\node[left=-1pt of predimg] {\small Predict};
\node[draw, rectangle, inner sep=1.2pt] (trueimg) at (2.15,-1.05) {
    \includegraphics[width=0.155\columnwidth]{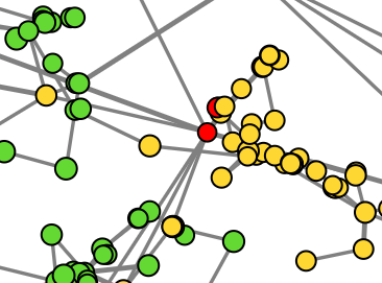}
};
\node[above=-2pt of trueimg] {\small True};
\draw[-{Stealth}, thick, draw=red] (predimg) -- (target);
\draw[-{Stealth}, thick, draw=red] (trueimg) -- (target);
\end{tikzpicture}%
\vspace{-0.8\baselineskip}%
\caption{Representative French-system activity map. The background shows the ground-truth classes, while the event region is enlarged to visualize the predicted and true local classifications.}
\label{fig:french_pred_true_graph}
\end{figure}
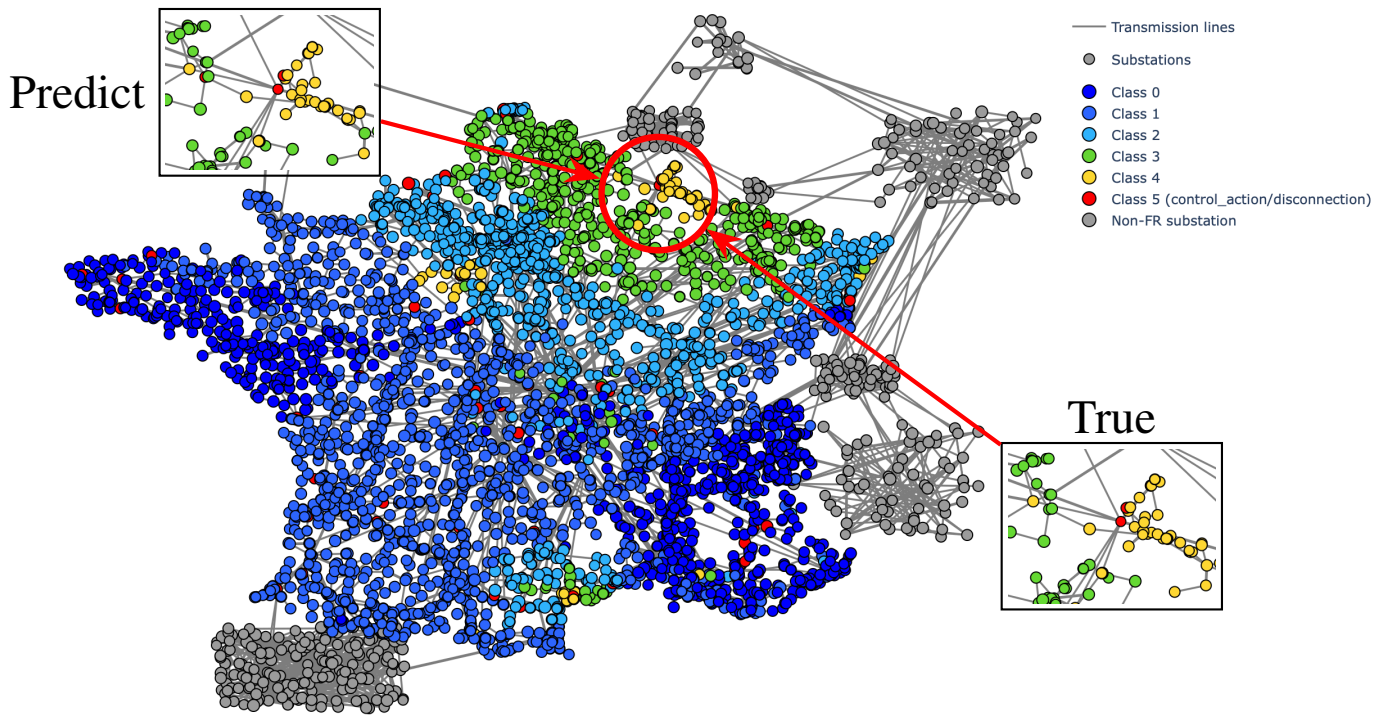

Fig.~\ref{fig:french_pred_true_graph} shows a representative contingency: the background is the ground-truth map, and the event region is enlarged for the predicted and true local classes.

\subsection{Adaptive Simplification and Real-Time Performance}
\label{sec:results_ams}

On the French test set ($6\,768$ scenarios), four models are compared: Full (node-breaker), Regional (three-hop heuristic of the transmission system operator), AMS oracle (true classes), and AMS Prediction (predicted classes). Table~\ref{tab:ams_variable_reduction} reports average system sizes for $\kappa\in\{1,\ldots,5\}$. Raising $\kappa$ shrinks the retained model but, for AMS Prediction, raises the risk of simplifying dynamically relevant substations; AMS oracle bounds what substation-level simplification can achieve with a perfect retain set.

\begin{figure*}[!t]
\centering
\includegraphics[width=\textwidth]{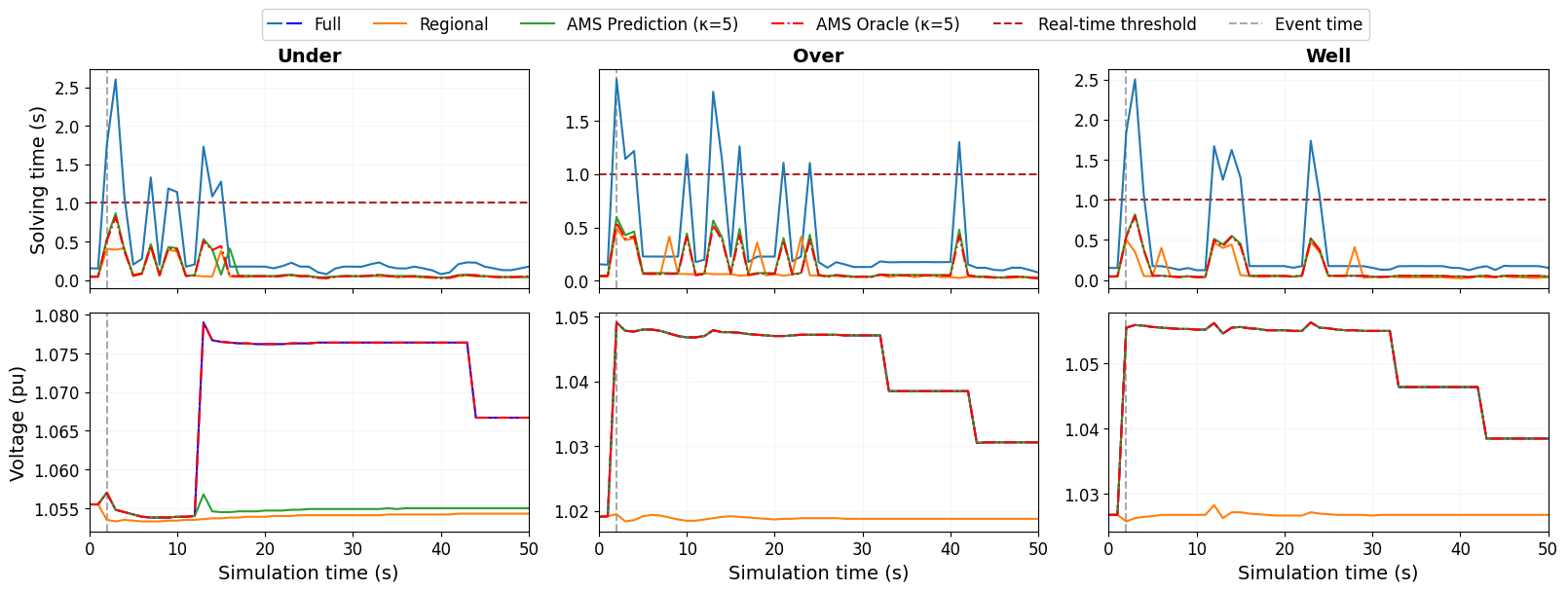}%
\vspace{-0.2\baselineskip}%
\makeatletter
\begingroup
\renewcommand{\@makecaption}[2]{%
  \@IEEEfigurecaptionsepspace
  \par\centering\normalfont\footnotesize {#1.}\nobreakspace\nobreakspace #2\par
}
\caption{Solving time (top) and voltage (bottom) for Under, Over, and Well ($\kappa{=}5$). Same 100\,s simulations as in Section~\ref{sec:methodology}, with initialization removed and time shifted so the contingency is at $t=2$\,s; plots shown until $50$\,s after settling.}
\label{fig:ams_timing_voltage}
\endgroup
\makeatother
\end{figure*}

Table~\ref{tab:ams_rt_dtw} illustrates the trade-off on three representative contingencies: Well (line), Over (bus), and Under (generator); Fig.~\ref{fig:ams_timing_voltage} shows the same cases. Solving time and voltage error follow Section~\ref{sec:ams_eval}. The underlying runs use the same 100\,s setup as in Section~\ref{sec:methodology}; for display, initialization is removed and the time axis is shifted so that the contingency appears at $t=2$\,s, with curves shown until $50$\,s after the response has settled. Well matches the oracle retain set; Over retains extra substations; Under removes a topology-dependent control action and incurs a non-negligible voltage error. Within a scenario the DTW RMSE stays constant across $\kappa$ while the critical retain set is preserved, and rises only when a needed substation is collapsed (Under, AMS Prediction for $\kappa\ge 3$).

\begin{table}
  \centering
  \caption{Average system size after substation-level simplification on the French test set (continuous, discrete, root functions).}
  \label{tab:ams_variable_reduction}
  \setlength{\tabcolsep}{1.5pt}
  \footnotesize
  \resizebox{\columnwidth}{!}{%
  \begin{tabular}{@{}r@{\hspace{0.4em}}c@{\hspace{0.55em}}c@{}}
    \toprule
    \textbf{System} & \textbf{System size} & \textbf{System size (\%)} \\
    \midrule
    Full & $(374\,989,\,397\,301,\,353\,160)$ & $(100.00\%,\,100.00\%,\,100.00\%)$ \\
    Regional & $(97\,760,\,130\,232,\,224\,645)$ & $(26.07\%,\,32.78\%,\,63.61\%)$ \\
    AMS oracle ($\kappa{=}1$) & $(249\,963,\,277\,215,\,295\,643)$ & $(66.66\%,\,69.77\%,\,83.71\%)$ \\
    AMS oracle ($\kappa{=}2$) & $(195\,427,\,224\,557,\,270\,209)$ & $(52.12\%,\,56.52\%,\,76.51\%)$ \\
    AMS oracle ($\kappa{=}3$) & $(147\,435,\,178\,201,\,247\,819)$ & $(39.32\%,\,44.85\%,\,70.17\%)$ \\
    AMS oracle ($\kappa{=}4$) & $(121\,578,\,153\,212,\,235\,745)$ & $(32.42\%,\,38.56\%,\,66.75\%)$ \\
    AMS oracle ($\kappa{=}5$) & $(101\,440,\,133\,574,\,226\,164)$ & $(27.05\%,\,33.62\%,\,64.04\%)$ \\
    AMS Prediction ($\kappa{=}1$) & $(273\,296,\,299\,774,\,306\,547)$ & $(72.88\%,\,75.45\%,\,86.80\%)$ \\
    AMS Prediction ($\kappa{=}2$) & $(208\,895,\,237\,556,\,276\,480)$ & $(55.71\%,\,59.79\%,\,78.29\%)$ \\
    AMS Prediction ($\kappa{=}3$) & $(155\,127,\,185\,647,\,251\,422)$ & $(41.37\%,\,46.73\%,\,71.19\%)$ \\
    AMS Prediction ($\kappa{=}4$) & $(128\,190,\,159\,546,\,238\,779)$ & $(34.19\%,\,40.16\%,\,67.61\%)$ \\
    AMS Prediction ($\kappa{=}5$) & $(104\,033,\,135\,996,\,227\,294)$ & $(27.74\%,\,34.23\%,\,64.36\%)$ \\
    \bottomrule
  \end{tabular}%
  }
\end{table}

\begin{table}[!t]
  \centering
  \caption{Real-time performance and voltage fidelity across scenarios. Speedup is the Full-to-simplified $\lVert T_{\mathrm{sol}}\rVert_{\infty}$ ratio.}
  \label{tab:ams_rt_dtw}
  \setlength{\tabcolsep}{2.5pt}
  \footnotesize
  \resizebox{\columnwidth}{!}{%
  \begin{tabular}{llcccc}
    \toprule
    \textbf{Scenario} & \textbf{System} & \textbf{RT Viol.} & $\lVert T_{\mathrm{sol}}\rVert_{\infty}$ & \textbf{DTW RMSE} & \textbf{Speedup} \\
    \midrule
    \multirow{12}{*}{Well (Line)}
      & Full & 9 & 2.4999 & ref & ref \\
      & Regional & 0 & 0.5032 & 0.018 & 4.97x \\
      & AMS oracle ($\kappa{=}1$) & 5 & 1.7470 & $2.89\times10^{-5}$ & 1.43x \\
      & AMS oracle ($\kappa{=}2$) & 1 & 1.1905 & $2.89\times10^{-5}$ & 2.10x \\
      & AMS oracle ($\kappa{=}3$) & 0 & 0.9867 & $2.89\times10^{-5}$ & 2.53x \\
      & AMS oracle ($\kappa{=}4$) & 0 & 0.8871 & $2.89\times10^{-5}$ & 2.82x \\
      & AMS oracle ($\kappa{=}5$) & 0 & 0.7983 & $2.89\times10^{-5}$ & 3.13x \\
      & AMS Prediction ($\kappa{=}1$) & 5 & 1.7215 & $2.89\times10^{-5}$ & 1.45x \\
      & AMS Prediction ($\kappa{=}2$) & 1 & 1.1490 & $2.89\times10^{-5}$ & 2.18x \\
      & AMS Prediction ($\kappa{=}3$) & 1 & 1.0074 & $2.89\times10^{-5}$ & 2.48x \\
      & AMS Prediction ($\kappa{=}4$) & 0 & 0.9207 & $2.89\times10^{-5}$ & 2.72x \\
      & AMS Prediction ($\kappa{=}5$) & 0 & 0.8179 & $2.89\times10^{-5}$ & 3.06x \\
    \midrule
    \multirow{12}{*}{Over (Bus)}
      & Full & 10 & 1.8891 & ref & ref \\
      & Regional & 0 & 0.4907 & 0.018 & 3.85x \\
      & AMS oracle ($\kappa{=}1$) & 2 & 1.3188 & $1.07\times10^{-5}$ & 1.43x \\
      & AMS oracle ($\kappa{=}2$) & 0 & 0.9473 & $1.07\times10^{-5}$ & 1.99x \\
      & AMS oracle ($\kappa{=}3$) & 0 & 0.6961 & $1.07\times10^{-5}$ & 2.71x \\
      & AMS oracle ($\kappa{=}4$) & 0 & 0.5851 & $1.07\times10^{-5}$ & 3.23x \\
      & AMS oracle ($\kappa{=}5$) & 0 & 0.5473 & $1.07\times10^{-5}$ & 3.45x \\
      & AMS Prediction ($\kappa{=}1$) & 6 & 1.5622 & $1.07\times10^{-5}$ & 1.21x \\
      & AMS Prediction ($\kappa{=}2$) & 2 & 1.2446 & $1.07\times10^{-5}$ & 1.52x \\
      & AMS Prediction ($\kappa{=}3$) & 1 & 1.0397 & $1.07\times10^{-5}$ & 1.82x \\
      & AMS Prediction ($\kappa{=}4$) & 0 & 0.9051 & $1.07\times10^{-5}$ & 2.09x \\
      & AMS Prediction ($\kappa{=}5$) & 0 & 0.5978 & $1.07\times10^{-5}$ & 3.16x \\
    \midrule
    \multirow{12}{*}{Under (Generator)}
      & Full & 9 & 2.6017 & ref & ref \\
      & Regional & 0 & 0.4257 & 0.019 & 6.11x \\
      & AMS oracle ($\kappa{=}1$) & 7 & 2.3077 & $1.45\times10^{-5}$ & 1.13x \\
      & AMS oracle ($\kappa{=}2$) & 3 & 1.7373 & $1.45\times10^{-5}$ & 1.50x \\
      & AMS oracle ($\kappa{=}3$) & 1 & 1.0800 & $1.45\times10^{-5}$ & 2.41x \\
      & AMS oracle ($\kappa{=}4$) & 0 & 0.8831 & $1.45\times10^{-5}$ & 2.95x \\
      & AMS oracle ($\kappa{=}5$) & 0 & 0.8199 & $1.45\times10^{-5}$ & 3.17x \\
      & AMS Prediction ($\kappa{=}1$) & 8 & 2.3235 & $1.45\times10^{-5}$ & 1.12x \\
      & AMS Prediction ($\kappa{=}2$) & 2 & 1.5278 & $1.45\times10^{-5}$ & 1.70x \\
      & AMS Prediction ($\kappa{=}3$) & 1 & 1.1806 & 0.014 & 2.20x \\
      & AMS Prediction ($\kappa{=}4$) & 0 & 0.9478 & 0.014 & 2.75x \\
      & AMS Prediction ($\kappa{=}5$) & 0 & 0.8672 & 0.014 & 3.00x \\
    \bottomrule
  \end{tabular}%
  }
\end{table}

\section{Discussion}
\label{sec:discussion}

Component-level dynamic activity can be predicted from the pre-disturbance operating point and contingency description alone when the labels remain physically meaningful and the decoder is conditioned on the specific target--contingency pair. The Nordic~32 and French experiments substantiate that claim, from cut construction and architecture choice through industrial-scale prediction and AMS. A purely heuristic cut ladder is insufficient. In Table~\ref{tab:nordic_cut_candidates} and Fig.~\ref{fig:cut_pareto_voltage}, candidate~A assigns most active samples to class~4 and is hard to learn, whereas candidate~C balances class shares while remaining consistent with morphological severity $\varphi$. Anchoring $\tau_0=10^{-7}$ to solver precision makes class~$0$ numerical inactivity rather than a mild physical response (Fig.~\ref{fig:delta_class_examples}). The same objectives yield tighter French cuts (Table~\ref{tab:kpi_cuts}); the inactive dominance of the French datasets in Fig.~\ref{fig:dataset_class_dist} therefore reflects localized post-event dynamics rather than a deficiency of the cut procedure.

On Nordic~32, Optuna selects candidate~C in Table~\ref{tab:nordic_hpo}, with the highest validation scores on both tasks. Candidate~A shares the same hidden width as~C but uses a much smaller learning rate and reaches the epoch budget with lower scores; candidate~B is wider and uses larger identity and pair dimensions, yet still trails~C. Among these three trials, neither a wider backbone nor larger embedding dimensions alone yield the best validation score.

The architecture comparison in Table~\ref{tab:nordic_arch_test} isolates the contribution of pair context and of graph structure. Conditioning the decoder on the target--contingency pair already raises accuracy relative to the local XGBoost baseline, and introducing graph embeddings yields a further, consistent gain. Among the four edge-aware encoders balanced accuracy differs by at most $0.006$, while GINE has the fewest parameters and the lowest median inference time; a heavier encoder is therefore not justified for the intended real-time use.

The same pipeline extends to the French transmission model, where $\mathrm{Acc}_{1}\approx 0.92$--$0.93$ despite stronger inactive dominance and a much larger graph (Table~\ref{tab:french_arch_test}). Missed classes usually stay neighboring. For AMS screening, a one-class overprediction only retains an extra substation and is therefore conservative, whereas a one-class underprediction at the retain/collapse boundary can drop a substation that still matters. Exact and scenario-level scores are lower than on Nordic~32, partly because secondary voltage control couples substations through a telecommunication layer that the present graphs omit. Class~5 (flag) sites can therefore appear in otherwise inactive electrical neighborhoods (Fig.~\ref{fig:french_pred_true_graph}) with no corresponding path in the learned representation. The protocol scales without redesign; adding telecommunication/control topology is the next step for those long-range couplings.

Deployed for AMS, the predicted map is judged against both Full and Regional (Table~\ref{tab:ams_rt_dtw} and Fig.~\ref{fig:ams_timing_voltage}). Regional is faster, but the voltage DTW is larger ($\sim10^{-2}$) and the retain rule is static: the three-hop neighborhood around the event substation does not follow the components that respond. AMS oracle shows that contingency-specific retain sets can meet the real-time budget with margin while keeping voltage DTW at $\sim10^{-5}$\,pu, below the solver tolerance of $10^{-4}$\,pu, and thereby sets the limit of substation-level simplification when the retain set is perfect. AMS Prediction approaches that oracle whenever the critical retain set is recovered (Well and Over). The residual risk is underprediction of substations that host topology-dependent control actions: simplifying such a substation changes the simulated control response and produces a non-negligible voltage error (Under). Overprediction retains extra substations, preserves fidelity, and only reduces the speedup (Over). A modest excess of overprediction is therefore acceptable if it suppresses underprediction of control-critical substations, because the cost of overprediction is computational, whereas the cost of underprediction is trajectory error.

\section{Conclusion}
\label{sec:conclusion}

This paper presented a pair-aware GINE framework with two task-specific models for predicting component-level dynamic activity from the pre-disturbance operating point and contingency description. Physically aware KPI cuts define the ordinal targets, and the decoder is conditioned on the target--contingency pair. On Nordic~32, held-out balanced accuracy rises from $0.859$ / $0.883$ (XGBoost) to $0.893$ / $0.921$ with pair-aware MLP fusion, and to about $0.91$--$0.94$ with edge-aware graph encoders; GINE attains $0.914$ / $0.934$ at a median inference cost of about $1.4$\,ms per scenario, below NNConv at about $2.05$\,ms. On the French system ($>$6\,000 buses), within-one accuracy is $0.930$ (voltage) and $0.917$ ($S_{\text{power}}$), with exact accuracy $0.679$ and $0.792$.

Used for AMS on $6\,768$ French test scenarios, the predicted maps at $\kappa{=}5$ cut average continuous and discrete model size by about $72\%$ and $66\%$. On representative contingencies, the same threshold yields speedups of $3.04$x--$3.16$x with voltage error below $10^{-4}$\,pu when critical substations are retained; underprediction of topology-dependent control sites remains the main residual risk. AMS is thus one demonstrated downstream use of the activity map.

Future work will investigate alternative definitions of the electrical-coupling feature and their effect on prediction performance, introduce additional targets such as system- or component-level frequency dynamic activity, and evaluate the methodology for additional applications, particularly online DSA and N-1 contingency screening and prioritization.

\section*{Acknowledgment}

\marktext{This work has received funding from the European Commission under the Horizon Europe programme, Grant Agreements No.~101136119 (TwinEU project) and No.~101120278 (DENSE). Additional support was provided through the TRAISIM project, which is funded by CRESYM.}

\balance
\bibliographystyle{IEEEtran}
\bibliography{biblio}

@inproceedings{xu2019howpowerful,
  author    = {Xu, Keyulu and Hu, Weihua and Leskovec, Jure and Jegelka, Stefanie},
  title     = {How Powerful Are Graph Neural Networks?},
  booktitle = {International Conference on Learning Representations},
  year      = {2019}
}

@inproceedings{hu2020strategies,
  author    = {Hu, Weihua and Liu, Bowen and Gomes, Joseph and Zitnik, Marinka and Liang, Percy and Pande, Vijay and Leskovec, Jure},
  title     = {Strategies for Pre-Training Graph Neural Networks},
  booktitle = {International Conference on Learning Representations},
  year      = {2020}
}

@inproceedings{varbella2024powergraph,
  author    = {Varbella, Anna and Amara, Kenza and Gjorgiev, Blazhe and El-Assady, Mennatallah and Sansavini, Giovanni},
  title     = {{PowerGraph}: A Power Grid Benchmark Dataset for Graph Neural Networks},
  booktitle = {Advances in Neural Information Processing Systems},
  year      = {2024}
}

@article{kazim2025edgegnn,
  author  = {Kazim, Muhammad and Pirim, Harun and Le, Chau and Le, Trung and Yadav, Om Prakash},
  title   = {Edge-Level Explainable Graph Neural Networks with Network Centric Features for Transmission Line Failure Prediction in Power Grids},
  journal = {Sustainable Energy, Grids and Networks},
  volume  = {44},
  pages   = {101969},
  year    = {2025},
  doi     = {10.1016/j.segan.2025.101969}
}

@ARTICLE{dispatcher_bucciero_1991,
  author={Bucciero, J.M. and Dodge, J.A. and Gillespie, R.N. and Hinkel, R.O. and Mann, S.L. and Martin, S.M. and Schulte, R.P. and Hayward, D.},
  journal={IEEE Transactions on Power Systems}, 
  title={Dispatcher training simulators: lessons learned}, 
  year={1991},
  volume={6},
  number={2},
  pages={594-604},
  doi={10.1109/59.76702}
}

@inproceedings{uses_chu_1991,
    author = {Chu, R. F. and Dobrowolski, E. J. and Barr, E. J. and McGeehan, J. and Scheurer, D. and Nodehi, K.},
    title = {The Uses of an Operator Training Simulator for System Restoration},
    booktitle = {Proc. IEEE Power Industry Computer Application Conference (PICA)},
    year = {1991},
    pages = {171--177},
    doi={10.1109/pica.1991.160625}
}

@article{experiences_miller_1993,
    author = {Miller, G. and Storey, A. B. and Vadari, S. V. and Brewer, K.},
    title = {Experiences Using the Dispatcher Training Simulator as a Training Tool},
    journal = {IEEE Transactions on Power Systems},
    volume = {8},
    number = {3},
    pages = {1126--1132},
    year = {1993},
    doi={10.1109/59.260885}
}

@article{support_gomez_2011,
  title     = {Support Vector Machine-Based Algorithm for Post-Fault Transient Stability Status Prediction Using Synchronized Measurements},
  author    = {Gomez, F. and Rajapakse, A. and Annakkage, Udaya and Fernando, Ioni T.},
  journal   = {IEEE Transactions on Power Systems},
  year      = {2011},
  doi       = {10.1109/tpwrs.2010.2082575},
  volume = {26},
  number = {3},
  pages = {1474--1483}
}

@article{rotor_sun_2024,
  title     = {Rotor Angle Stability Prediction Using Temporal and Topological Embedding Deep Neural Network Based on Grid-informed Adjacency Matrix},
  author    = {Sun, Peiyuan and Huo, Long and Chen, Xin and Liang, Siyuan},
  journal   = {Journal of Modern Power Systems and Clean Energy},
  year      = {2024},
  doi       = {10.35833/mpce.2023.000364},
  volume = {12},
  number = {3},
  pages = {695--706}
}

@article{grubased_okasha_2026,
  title     = {A {GRU}-Based Model for Real-Time Transient Stability Prediction in Electrical Power Systems},
  author    = {Okasha, Aliaa A. and Mansour, Diaa-Eldin A. and Zaky, Ahmed B. and Nguyen, Dinh Hoa and Megahed, Tamer F.},
  journal   = {Lecture Notes in Electrical Engineering},
  year      = {2026},
  doi       = {10.1007/978-981-95-5941-1_14},
  pages = {179--188}
}

@inproceedings{power_an_2019,
  title     = {Power system transient stability margin assessment using steady-state information},
  author    = {An, Jun and Mu, Gang and Li, Zonghan and Liu, Daowei},
  booktitle   = {Proc. IEEE Innovative Smart Grid Technologies - Asia (ISGT Asia)},
  year      = {2019},
  doi       = {10.1109/isgt-asia.2019.8881332},
  pages = {642--647}
}

@article{recurrent_huang_2020,
  title     = {Recurrent Graph Convolutional Network-Based Multi-Task Transient Stability Assessment Framework in Power System},
  author    = {Huang, Jiyu and Guan, Lin and Su, Yinsheng and Yao, Haicheng and Guo, Mengxuan and Zhong, Zhi},
  journal   = {IEEE Access},
  year      = {2020},
  doi       = {10.1109/access.2020.2991263},
  volume = {8},
  pages = {93283--93296}
}

@inproceedings{measurementbased_chai_2016,
  title     = {Measurement-based system reduction using autoregressive model},
  author    = {Chai, Jidong and Liu, Yong and Liu, Yilu and Bhatt, N. and {Del Rosso}, A. and Farantatos, E.},
  booktitle   = {Proc. IEEE/PES Transmission and Distribution Conference and Exposition (T\&D)},
  year      = {2016},
  doi       = {10.1109/tdc.2016.7519981},
  pages = {1--5}
}

@inproceedings{online_jiang_2021,
  title     = {Online Tuning of Dynamic Equivalents for Large-Scale Power Systems Using Wide-area Measurements},
  author    = {Jiang, Zhihao and Tong, Ning and Zhu, Lin and Liu, Yilu},
  booktitle   = {Proc. IEEE Power \& Energy Society General Meeting (PESGM)},
  year      = {2021},
  doi       = {10.1109/pesgm46819.2021.9637846},
  pages = {1--5}
}

@inproceedings{parallel_aristidou_2014,
  title     = {Parallel Computing and Localization Techniques for Faster Power System Dynamic Simulations},
  booktitle = {Proceedings of Cigre 2014 Belgium Conference},
  author    = {Aristidou, Petros and {Van Cutsem}, Thierry},
  year      = {2014}
}

@article{deb2002nsga2,
  author    = {Deb, Kalyanmoy and Pratap, Amrit and Agarwal, Sameer and Meyarivan, T.},
  title     = {A Fast and Elitist Multiobjective Genetic Algorithm: {NSGA-II}},
  journal   = {IEEE Transactions on Evolutionary Computation},
  year      = {2002},
  volume    = {6},
  number    = {2},
  pages     = {182--197},
  doi       = {10.1109/4235.996017}
}

@article{marquez2020emodits,
  author    = {M{\'a}rquez-Grajales, Aldo and Acosta-Mesa, H{\'e}ctor-Gabriel and Mezura-Montes, Efr{\'e}n and Graff, Mario},
  title     = {A Multi-Breakpoints Approach for Symbolic Discretization of Time Series},
  journal   = {Knowledge and Information Systems},
  year      = {2020},
  volume    = {62},
  number    = {7},
  pages     = {2795--2834},
  doi       = {10.1007/s10115-020-01437-4}
}

@article{PSCC_2026,
	title = {Towards an open-source real-time operator-training platform: Analysis of computational efficiency},
	doi = {10.1016/j.epsr.2026.113626},
	author = {Panagi, Savvas and Vermeulen, Thibaut and Joseph, Geethu and Aristidou, Petros},
	journal = {Electric Power Systems Research},
	year = {2026},
	litmapsId = {316563918},
  volume = {262},
  pages = {113626}
}

@inproceedings{timevariant_xia_2022,
	title = {Time-variant Nonlinear Participation Factors Considering Resonances in Power Systems},
	doi = {10.1109/pesgm48719.2022.9916932},
	author = {Xia, Tianwei and Sun, Kai},
	booktitle = {Proc. IEEE Power \& Energy Society General Meeting (PESGM)},
	year = {2022},
	litmapsId = {233785023},
  pages = {1--5}
}

@misc{dynawo,
  title        = {Dynawo: A Comprehensive Power System Dynamic Simulator},
  note         = {Supported by the Linux Foundation Energy initiative},
  year         = {2025},
  url          = {https://dynawo.github.io}
}

@article{hagmar_dvsm_2020,
  author  = {Hagmar, Hannes and Eriksson, Robert and Le, Anh Tuan},
  title   = {Fast Dynamic Voltage Security Margin Estimation: Concept and Development},
  journal = {IET Smart Grid},
  volume  = {3},
  number  = {4},
  pages   = {470--478},
  year    = {2020},
  doi     = {10.1049/iet-stg.2019.0278}
}

@article{best_west_2021,
	title = {Best practice in statistics: The use of log transformation},
	doi = {10.1177/00045632211050531},
	author = {West, R.},
	journal = {Annals of Clinical Biochemistry},
	year = {2022},
	litmapsId = {276671743},
  volume = {59},
  number = {3},
  pages = {162--165}
}

@inproceedings{optuna_akiba_2019,
	title = {Optuna: A Next-generation Hyperparameter Optimization Framework},
	doi = {10.1145/3292500.3330701},
	author = {Akiba, Takuya and Sano, Shotaro and Yanase, Toshihiko and Ohta, Takeru and Koyama, Masanori},
	booktitle = {Proc. 25th ACM SIGKDD International Conference on Knowledge Discovery \& Data Mining},
	year = {2019},
	litmapsId = {257480659},
  pages = {2623--2631}
}

@inproceedings{model_chaniotis_2005,
	title = {Model reduction in power systems using {K}rylov subspace methods},
	doi = {10.1109/pes.2005.1489192},
	author = {Chaniotis, D. and Pai, M.A.},
	booktitle = {Proc. IEEE Power Engineering Society General Meeting},
	year = {2005},
	litmapsId = {53532828}
}

@article{identification_podmore_1978,
	title = {Identification of Coherent Generators for Dynamic Equivalents},
	doi = {10.1109/tpas.1978.354620},
	author = {Podmore, R.},
	journal = {IEEE Transactions on Power Apparatus and Systems},
	year = {1978},
	litmapsId = {192100055},
  volume = {PAS-97},
  number = {4},
  pages = {1344--1354}
}

@article{predicting_titz_2024,
  title = {Predicting dynamic stability from static features in power grid models using machine learning},
  author = {Titz, Maurizio and Kaiser, Franz and Kruse, Johannes and Witthaut, Dirk},
  journal = {Chaos: An Interdisciplinary Journal of Nonlinear Science},
  year = {2024},
  doi = {10.1063/5.0175372},
  volume = {34},
  number = {1}
}

@article{zhang_tsa_scrf_2026,
  title = {Power-System Transient Stability Assessment Based on High-Level Sample Feature Extraction and Model Updating},
  author = {Zhang, Shuolin and Yu, Yue and Tao, Ye and Xue, Lin},
  journal = {Processes},
  volume = {14},
  number = {15},
  pages = {2468},
  year = {2026},
  doi = {10.3390/pr14152468}
}

@inproceedings{huang_gtsa_2023,
  author    = {Huang, Jiyu and Guan, Lin and Cai, Zihan and Chen, Liukai and Chen, Haoying and Chen, Zhiying},
  title     = {A Global Attention Pooling-Based Graph Learning Scheme for Generator-Level Transient Stability Assessment},
  booktitle = {2023 IEEE Power \& Energy Society General Meeting (PESGM)},
  year      = {2023},
  pages     = {1--5},
  doi       = {10.1109/PESGM52003.2023.10253244}
}

@article{huang_gtsa_2025,
  author    = {Huang, Jiyu and Guan, Lin and Su, Yinsheng and Cai, Zihan and Chen, Liukai and Li, Yongzhe and Zhang, Jinyang},
  title     = {Generator-Level Transient Stability Assessment in Power System Based on Graph Deep Learning with Sparse Hybrid Pooling},
  journal   = {Electronics},
  year      = {2025},
  volume    = {14},
  number    = {6},
  pages     = {1180},
  doi       = {10.3390/electronics14061180}
}

@article{lu_screening_2024,
  author    = {Lu, Genghong and Bu, Siqi},
  title     = {Adaptive Stability Contingency Screening for Operational Planning Based on Domain-Adversarial Graph Neural Network},
  journal   = {IEEE Transactions on Power Systems},
  year      = {2024},
  volume    = {39},
  number    = {1},
  pages     = {1503--1516},
  doi       = {10.1109/TPWRS.2023.3262851}
}

@article{aththanayake_reduction_2024,
  author    = {Aththanayake, Lahiru and Hosseinzadeh, Nasser and Gargoom, Ameen and Alhelou, Hassan Haes},
  title     = {Power System Reduction Techniques for Planning and Stability Studies: A Review},
  journal   = {Electric Power Systems Research},
  year      = {2024},
  volume    = {227},
  pages     = {109917},
  doi       = {10.1016/j.epsr.2023.109917}
}

@article{salvador2007toward,
  title={Toward accurate dynamic time warping in linear time and space},
  author={Salvador, Stan and Chan, Philip},
  journal={Intelligent data analysis},
  volume={11},
  number={5},
  pages={561--580},
  year={2007},
  publisher={SAGE Publications Sage UK: London, England}
}

@inproceedings{chen2016xgboost,
  title={Xgboost: A scalable tree boosting system},
  author={Chen, Tianqi and Guestrin, Carlos},
  booktitle={Proceedings of the 22nd acm sigkdd international conference on knowledge discovery and data mining},
  pages={785--794},
  year={2016}
}

@article{brody2021attentive,
  title={How attentive are graph attention networks?},
  author={Brody, Shaked and Alon, Uri and Yahav, Eran},
  journal={arXiv preprint arXiv:2105.14491},
  year={2021}
}

@inproceedings{gilmer2017neural,
  title={Neural message passing for quantum chemistry},
  author={Gilmer, Justin and Schoenholz, Samuel S and Riley, Patrick F and Vinyals, Oriol and Dahl, George E},
  booktitle={International conference on machine learning},
  pages={1263--1272},
  year={2017},
  organization={Pmlr}
}

@article{shi2020masked,
  title={Masked label prediction: Unified message passing model for semi-supervised classification},
  author={Shi, Yunsheng and Huang, Zhengjie and Feng, Shikun and Zhong, Hui and Wang, Wenjin and Sun, Yu},
  journal={arXiv preprint arXiv:2009.03509},
  year={2020}
}

\end{document}